\documentclass[prl,twocolumn,english,superscriptaddress]{revtex4-1}

\usepackage[T1]{fontenc}
\usepackage[latin9]{inputenc}
\usepackage{color}
\usepackage{babel}
\usepackage{latexsym}
\usepackage{float}
\usepackage{amsmath}
\usepackage{graphicx}
\usepackage{times}   %% Times Roman font
\usepackage{esint}
\usepackage[unicode=true,pdfusetitle,
 bookmarks=false,colorlinks=true,citecolor=blue,urlcolor=blue,linkcolor=red]{hyperref}
\usepackage{multirow}
\makeatletter

\@ifundefined{textcolor}{}
{%
 \definecolor{BLACK}{gray}{0}
 \definecolor{WHITE}{gray}{1}
 \definecolor{RED}{rgb}{1,0,0}
 \definecolor{GREEN}{rgb}{0,1,0}
 \definecolor{BLUE}{rgb}{0,0,1}
 \definecolor{CYAN}{cmyk}{1,0,0,0}
 \definecolor{MAGENTA}{cmyk}{0,1,0,0}
 \definecolor{YELLOW}{cmyk}{0,0,1,0}
}

\@ifundefined{date}{}{\date{}}
\AtBeginDocument{
  
}
\makeatother

\newcommand{\SAVE}[1]{}

\newcommand{\prlsec}[1]{\emph{#1---}}

\newcommand{\ntc}{{Na$_2$Ti$_3$Cl$_8$ }}

\newcommand*{\angstrom}{\texttt{\normalfont{\AA}}}
\newcommand{\beginsupplement}{%
        \setcounter{table}{0}
        \renewcommand{\thetable}{S\arabic{table}}%
        \setcounter{figure}{0}
        \renewcommand{\thefigure}{S\arabic{figure}}%
     } 

\begin{document}
\renewcommand\abstractname{}

\title{Spin--lattice coupling and the emergence of the trimerized phase in the $S=1$ Kagome antiferromagnet Na$_2$Ti$_3$Cl$_8$}
\author{Arpita Paul}
\affiliation{Department of Chemical Engineering and Materials Science,University of Minnesota, Minneapolis, Minnesota 55455, USA}
\author{Chia-Min Chung}
\affiliation{Department of Physics and Arnold Sommerfeld Center for Theoretical Physics,
Ludwig-Maximilians-Universitat Munchen, Theresienstrasse 37, 80333 Munchen, Germany}
\author{Turan Birol}
\email{tbirol@umn.edu}
\affiliation{Department of Chemical Engineering and Materials Science,University of Minnesota, Minneapolis, Minnesota 55455, USA}
\author{Hitesh J. Changlani}
\email{hchanglani@fsu.edu}
\affiliation{Department of Physics, Florida State University, Tallahassee, Florida 32306, USA}
\affiliation{National High Magnetic Field Laboratory, Tallahassee, Florida 32304, USA}
\date{\today}

\begin{abstract}
Spin-1 antiferromagnets are abundant in nature, but few theories or results exist to understand their general properties and behavior, particularly in situations when geometric frustration is present. 
Here we study the $S=1$ Kagome compound \ntc using a combination of Density Functional Theory, Exact Diagonalization, and Density Matrix Renormalization Group methods to achieve a first principles supported explanation of exotic magnetic phases in this compound. 
We find that the effective magnetic Hamiltonian includes essential non-Heisenberg terms that do not stem from spin-orbit coupling, and both trimerized and spin-nematic magnetic phases are relevant. The experimentally observed structural transition to a breathing Kagome phase is driven by spin--lattice coupling, which favors the trimerized magnetic phase against the quadrupolar one. 
We thus show that lattice effects can be necessary to understand the magnetism in frustrated magnetic compounds, and surmise that \ntc is a compound which cannot be understood from only electronic or only lattice Hamiltonians, very much like VO$_2$. 
\end{abstract}

\maketitle
The search for exotic phases of matter in geometrical frustrated magnets has been an area of active research. 
To a large extent, effort has been focused on $S=1/2$ 2D materials~\cite{Nocera_Kagome_2007, YS_Lee_2012, Zorko_2017} which have seen a flurry of 
theoretical activity~\cite{White_kagome, Wen_Kagome, Depenbrock_Kagome, Iqbal_kagome, Norman_2016, Kumar_PRB16, Changlani_PRL18}. Less explored is the $S \geq 1$ case~\cite{Hida,Gotze}, 
where many candidate materials exist, but where the theoretical effort has not been proportionate to the experimental activity. 
This is partly based on the rationale that larger $S$ systems 
magnetically order at low temperature, however, there are many counter-examples to this intuition. For example, 
both theoretically and experimentally, it has been found that certain compounds do not conform to this scenario and instead form long-range 
non-magnetic states such as valence bond (simplex) or "trimerized" phases (in the case of the $S=1$ kagome~\cite{Arovas,Corboz,Changlani_PRB15,vondelft_PRB15,Ghosh_PRB16}). 
In some cases, a strongly quantum fluctuating phase or "spin liquid" is favored, as has been argued in the case of the nearly idealized Heisenberg 
$S=1$ pyrochlore~\cite{Plumb_NatPhys19, Zhang_PRL19, Iqbal_PRX19}), triangular lattices~\cite{Nakatsuji_2005,Balicas_2011,Serbyn_2011, Lauchli_Mila_Penc}, 
with second nearest neighbor and/or biquadratic couplings and possibly even the honeycomb lattice~\cite{Mahajan_honeycomb}. 
Further prohibiting deeper understanding of the physics of these materials is the interaction of magnetic degrees of freedom with the lattice, which provides an additional mechanism 
of relieving magnetic frustration. This work is thus motivated by the exploration of the interplay of magnetism with the lattice in $S=1$ kagome materials, 
which have multiple reported experimental realizations~\cite{Zhou_PRM18,Takagi_PRB17}. 

\begin{figure}
\centering
\includegraphics[width=\linewidth]{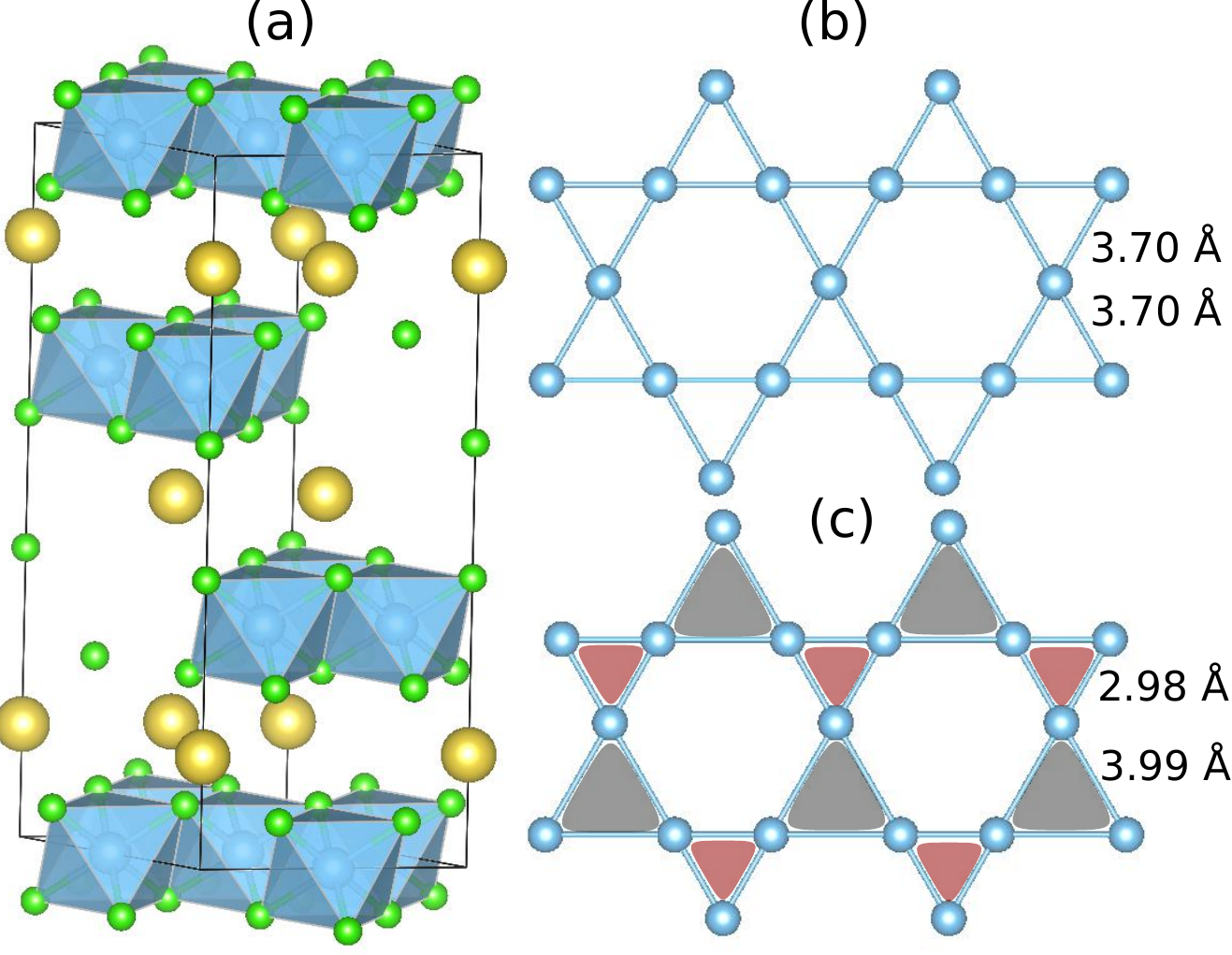}
	\caption{(Color online) (a) Crystal structure of Na$_2$Ti$_3$Cl$_8$ consists of layers of edge-sharing TiCl$_6$ octahedra, which are intercalated with Na ions. (b) At room temperature, Ti ions in each layer form ideal Kagome lattices (HT structure). (c) At low temperatures, a breathing distortion sets in, resulting in two different Ti-Ti bond lengths of 2.98~\AA and 3.99~\AA. 
}
\label{fig:fig1} 
\end{figure}	

\ntc, a compound that has been known for at least 24 years~\cite{Hinz1995}, 
has recently seen a resurgence of interest due to the underlying $S=1$ kagome physics, and its relevance to understanding 
the interplay between magnetic and lattice degrees of freedom \cite{Hanni_17, Kelly2019}. At room temperature, the compound has layers of titanium ions arranged in a kagome structure, as shown in Fig.~\ref{fig:fig1}. The titanium ions are in Ti$^{2+}$ configuration, so Hund's rules dictate a $3d^2$ configuration with $S=1$ magnetic moments. Experimentally, at low temperature (LT), \ntc has the "breathing kagome" or "trimerized" structure, 
referred to as in the literature as the $\gamma$ phase~\cite{Hanni_17} (Fig. \ref{fig:fig1}b). On heating the sample, at around $200$ K, a phase transition occurs 
to the undistorted kagome structure, the room temperature $\alpha$ phase~\cite{Hanni_17}, which we refer to as the high temperature (HT) phase. 
On cooling the sample from the HT phase, 
one reproducibly gets trapped in an ``intermediate" $\beta$ phase (IT phase)
which appears to be a distinct metastable state~\cite{Hanni_17}. Magnetic susceptibility drops sharply with decreasing temperature below the HT phase, consistent with $S=1$ atomic moments at HT phase, which are suppressed in the IT and LT phases as the crystal structure is trimerized \cite{Kelly2019, Hanni_17, Hinz1995}. 

Here we elucidate the magnetic ground state and explicate the mechanism of the breathing distortion in \ntc by a combination of first principles density functional theory (DFT), exact diagonalization (ED), and density matrix renormalization group (DMRG) approaches. We find that 
(i) the magnetic Hamiltonian that describes the interactions between atomic spins moments in the HT phase includes essential non-Heisenberg terms (biquadratic and ring-like exchange) that stem from higher order processes, and
(ii) due to the magnitude of these non-Heisenberg terms, the magnetic groundstate of the HT Hamiltonian is ferroquadrupolar (nematic) instead of trimerized. This implies that the breathing distortion of the lattice is necessary to stabilize the trimerized phase. 
We also find that (iii) the DFT calculations on the HT phase with Neel order point to no lattice instability, which implies that the trimerized ground state is stabilized through spin-lattice coupling. In other words, neither the lattice nor the magnetic Hamiltonians by themselves have any instabilities towards trimerization, but their combination gives rise to a coincident magnetic-structural transition. 

\prlsec{The Effective Hamiltonian}
Lack of information on the low-energy effective Hamiltonian is often a limiting factor in studies of frustrated magnetic materials. While there has been progress in downfolding approaches using 
quantum mechanical expectation values~\cite{Changlani_JCP2015, Zheng_FrontPhys2018}, 
here we adopt the classical fitting approach in conjunction with DFT that is now commonly used to extract magnetic Hamiltonians and parameter for real materials. 
(See, for example, Refs.~\cite{Spaldin,Jeschke_Valenti, Birol2018}.) We performed self-consistent DFT calculations for multiple magnetic configurations, including various collinear and non-collinear states, and extracted the final spin configurations and energies at the DFT level. We then fit the parameters of various magnetic models to these energies. 
%(We note that this approximation is rigorously valid only in the large $S$ limit where the spin moment is not significantly renormalized by quantum fluctuations.)

\begin{figure}
\centering
\includegraphics[width=\linewidth]{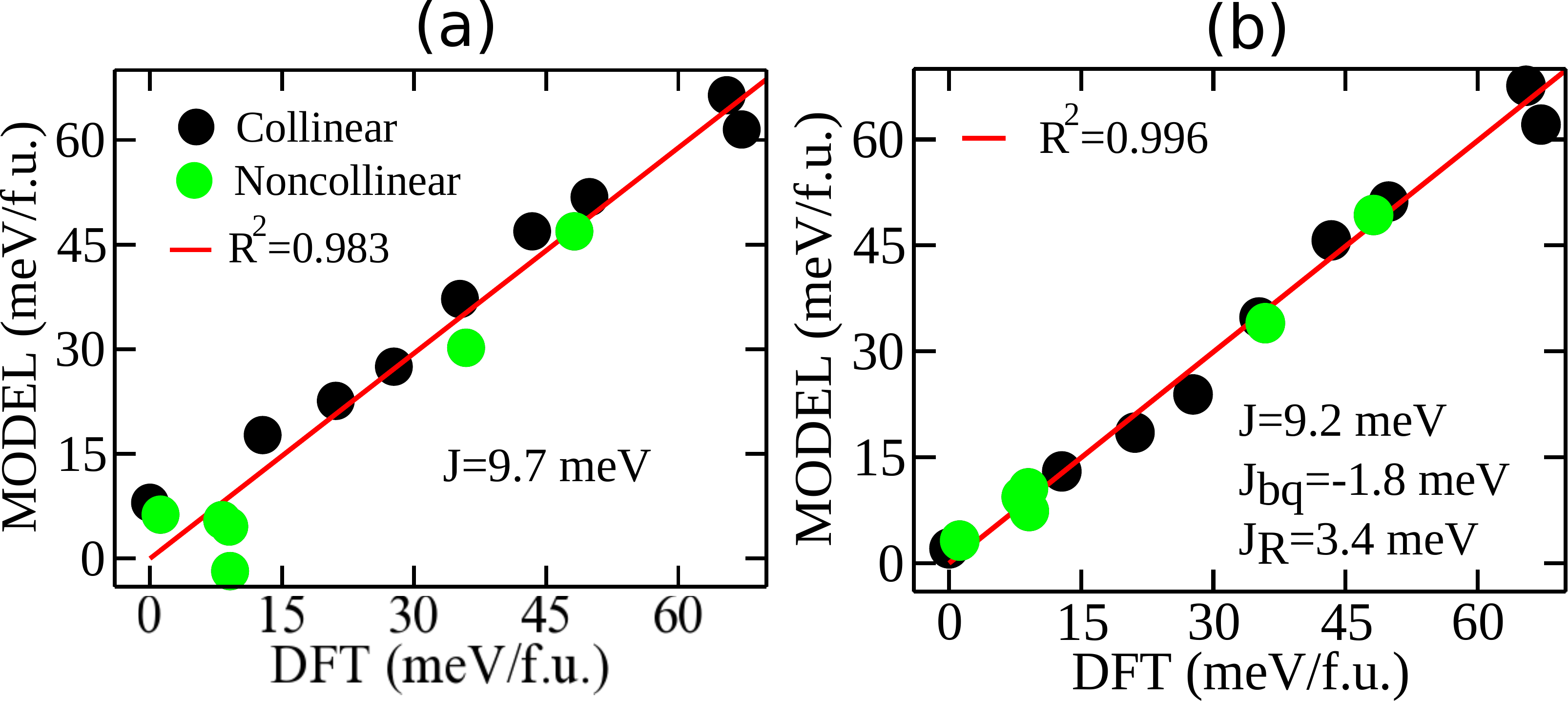}
\caption{(Color online) Fits of different effective model spin Hamiltonians to density functional theory data for $U=3$~eV. Each data point corresponds to a different magnetic configuration. The horizontal axis is the energy from the DFT calculation, and the vertical axis is the energy for the same configuration from the fitted model. (a) The fit to the model with only the nearest neighbor Heisenberg coupling. The energies of many non-collinear states are not reproduced well by the model.  (b) The model with biquadratic and ring-exchange couplings. The agreement is enhanced, with no clear outliers in the data. }
\label{fig:fig2} 
\end{figure}	

\begin{figure*}
\centering
\includegraphics[width=1\linewidth]{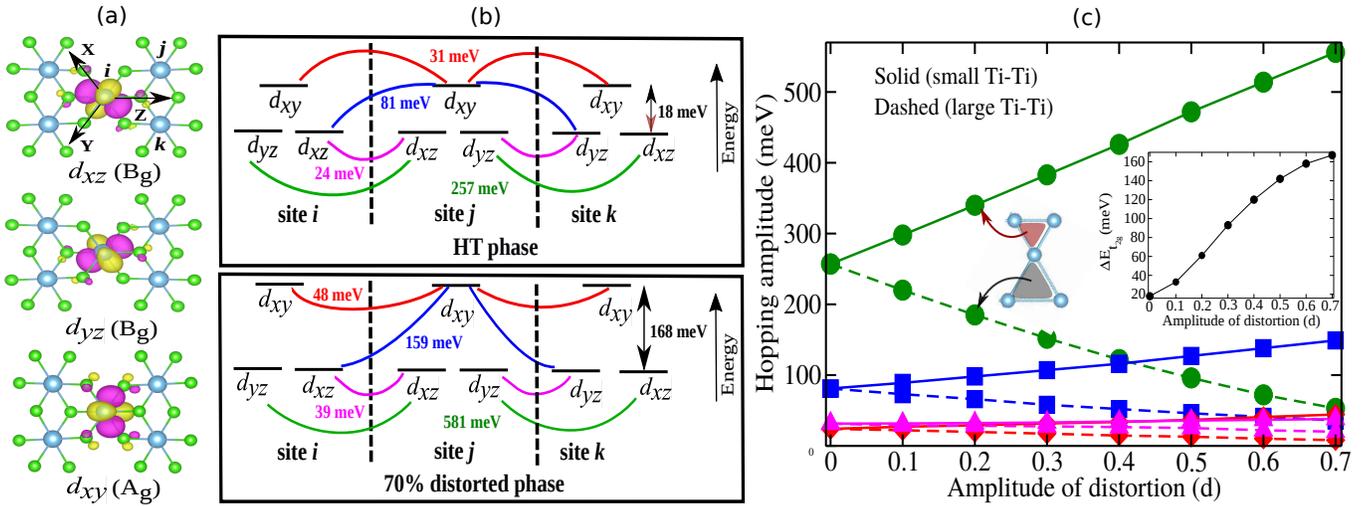}
	\caption{(Color online) (a) The $t_{2g}$-like Wannier functions obtained in the HT structure from nonmagnetic DFT calculations. While there is some hybridization with the Cl ions as expected, the Wannier orbitals have atomic character. (This is no longer the case in the LT structure, see the supplement for further details \cite{supplement}.) 
	(b) Sketch of the three $t_{2g}$ orbitals on a triangle and the hoppings between them in the HT phase (top), and in a hypothetical structure that is obtained by linearly interpolating the structural parameters (lattice constants and atomic positions) between the HT and LT phases. The values in the sketch are for the 70\% distorted structure, where 100\% distortion would correspond to the LT phase. 
	(c) The hopping amplitudes between the orbitals on neighboring atoms and the splitting between the orbitals on the same atom as a function of distortion amplitude. }
\label{fig:fig3} 
\end{figure*}

In Fig. \ref{fig:fig2}, we present the results of our DFT calculations for the HT structure, performed using the PBEsol exchange correlation functional with the on-site +U correction with $U=3$~eV \cite{VASP1, VASP2, PBEsol, Dudarev1998, supplement}. A fit to a nearest-neighbor only Heisenberg Hamiltonian captures the main trend of the energy with an antiferromagnetic nearest neighbor coupling; but the agreement is far from perfect, and  especially the non-collinear spin configurations' energy are not properly captured by the model (Fig.~\ref{fig:fig2}a). Possibly the simplest extension of the Hamiltonian is the biquadratic term $\sim \left( S_i \cdot S_j \right)^2$~\cite{Blume_Hsieh}. This biquadratic exchange is allowed by symmetry, and emerges in various spin-1 models due to higher order ($\sim t^4$, where $t$ is the hopping amplitude) perturbations which 
correspond to multiple electrons between two atoms~\cite{Bhatt_Yang, Mila_Zhang, fazekas1999, supplement}. At the same order in nearest neighbor hopping $t$, there also exists a ring exchange on the triangles with the form $\sim \left( S_i \cdot S_j \right)\left( S_i \cdot S_k \right)$. We include both of these terms to get the Hamiltonian  
\begin{multline}
  \mathcal{H}=J \sum_{\langle ij\rangle}{\mathbf{S}_{i}\cdot{\mathbf{S}_{j}}} + J_{bq}\sum_{\langle ij \rangle} \left({\mathbf{S}_{i}} \cdot{\mathbf{S}_{j}} \right)^{2} \\
  + \frac{J_{R}}{2} \sum_{\Delta=i,j,k} \left( \left( \mathbf{S}_i \cdot \mathbf{S}_j \right) \left(\mathbf{S}_i \cdot \mathbf{S}_k \right) + \left(\mathbf{S}_i \cdot \mathbf{S}_k \right) \left(\mathbf{S}_i \cdot \mathbf{S}_j \right) \right)
\label{eq:hamiltonian}
\end{multline}
where $\langle ij\rangle$ refers to nearest neighbor pairs and $J > 0$ is the Heisenberg coupling. 
The symmetrization in the ring exchange term is required to maintain Hermiticity of the Hamiltonian. 
Ring exchanges similar to this one have been proposed and studied in square lattices before \cite{Desai_Kaul}, but to the best of our knowledge, this form of the Hamiltonian has 
not been considered for a Kagome system before. The inclusion of more terms make the fit better, as expected (Fig.~\ref{fig:fig2}). We find that while the nearest neighbor antiferromagnetic Heisenberg coupling is the strongest term, both $J_{bq}$ and $J_R$ are nonzero and significant. In the supplementary information \cite{supplement} we provide a jackknife analysis to show that the data is not over-fit, and discuss the possibility of other Hamiltonians that can be fit to the DFT data but require further neighbor hopping terms. 

Wannier analysis of the electronic structure of Na$_2$Ti$_3$Cl$_8$ provides insight into the reason that the Hamiltonian attains this complicated form, and also to how the $J$ coefficients behave under the structural transition. In Fig. \ref{fig:fig3}a, we show the t$_\textrm{2g}$-like Wannier functions on the Ti atoms. The Ti cations are at Wyckoff position 9e with site symmetry 2/m (C$_\textrm{2h}$). This low symmetry of the crystal field further splits the 3 t$_\textrm{2g}$ orbitals into $t_{2g}\rightarrow A_g+B_g+B_g$, but our first principles calculations indicate that the two $B_g$ orbitals ($xz$ and $yz$) are degenerate within numerical noise, and only the $A_g$ ($xy$) orbital has a different energy. In Fig.~\ref{fig:fig3}b, we show the hoppings between the 3 t$_\textrm{2g}$-like orbitals in the HT phase. There are at least 3 different $t$ values that are large and hence contribute significantly to the exchange processes. While we do not attempt to solve this model explicitly, we note that it is rich enough to give rise to the biquadratic exchange. To derive a biquadratic term starting from an orbital model, Ref.~\onlinecite{Mila_Zhang} considered a model with 2 electrons on 3 orbitals, whereas Ref.~\onlinecite{fazekas1999} used a two orbital model with same-orbital hoppings. The Wannier model for Na$_2$Ti$_3$Cl$_8$ includes both of these terms and hence it is no surprise that a biquadratic term emerges. The ring exchange term $\sim \left( S_i \cdot S_j \right)\left( S_i \cdot S_k \right)$ can emerge from simultaneous hopping of two electrons from site $i$ to sites $j$ and $k$. Given that the largest hopping element (257~meV) is between alternating $xz$ and $yz$ orbitals in nearest neighbor atoms, this term will be proportional to the highest $t^4$ factor, and is expected to be significant as well. 

Due to the low symmetry and very small Ti--Ti distances, it is technically challenging to stabilize many different magnetic states and calculate $J$'s with high precision in the LT phase using DFT. Instead, in Fig.~\ref{fig:fig3}c, we present the evolution of the hopping parameters and the $A_g$--$B_g$ splitting, obtained from the Wannier tight binding models, as the crystal structure is linearly interpolated between the HT and the LT phases. As the crystal structure gets close to that of the LT phase, the Ti--Ti hoppings only in the larger triangles all go to zero. Two trends are evident: 1) The $A_g$--$B_g$ energy separation increases by almost an order of magnitude in the LT phase. 2) The largest hopping amplitude (shown in green) becomes even larger compared to all the other terms in the LT phase. Thus, in the low temperature phase, the antiferromagnetic exchange $J$ is enhanced because of the increasing $A_g$--$B_g$ separation makes the system an effectively half-filled system with increasing $t$. Also, $J_{bq}/J$ is suppressed, since a model with two half-filled orbitals per atom with significant hopping only between a pair of them cannot have biquadratic exchange according to the models proposed so far \cite{Bhatt_Yang, Mila_Zhang, fazekas1999}. $J_{R}/J$, on the other hand, is not easy to predict, since there are multiple processes that contribute to this term, some of which (e.g. the one that involves hopping from $xz$ or $yz$ orbitals to $xy$ orbital) are suppressed, whereas some of which (e.g. the one that involves hopping between $xz$ and $yz$ orbitals) are enhanced. 
%%%%%%%%%%%%%%%%%%%%%%%%%%%%%%%%%%%%%%%%%%%%%%%%%%%%%%%

\prlsec{Exact diagonalization and density matrix renormalization group}
DFT calculations provide estimates of the parameters of the effective Hamiltonian, but they
do not conclusively tell us the nature of the quantum many-body ground state. Thus we appeal to and extend our results based on 
previous numerical and analytic calculations on the idealized $S=1$ kagome system which has been previously studied with a variety of 
approaches~\cite{Changlani_PRB15, vondelft_PRB15, Oitmaa_Singh, Ghosh_PRB16}. For positive biquadratic interactions, 
the existence of a trimerized state was established~\cite{Arovas,Corboz}. 
This state was found to persist to negative biquadratic interactions $J_{bq}/J \approx -0.16$ below which it transitions to a spin nematic- 
a state with ferroquadrupolar order~\cite{Changlani_PRB15}. 

\begin{figure}
\includegraphics[width=0.9\linewidth]{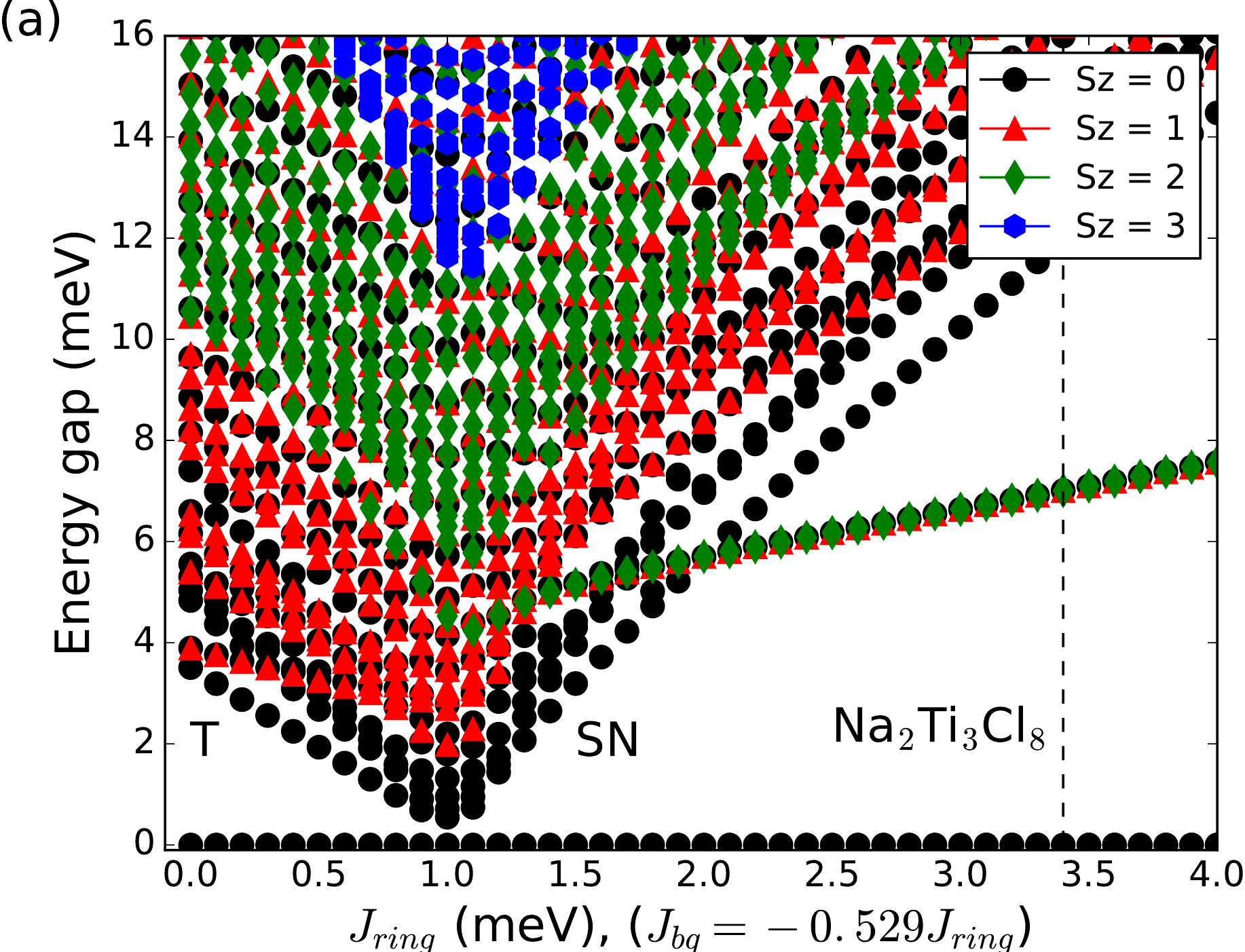}
\includegraphics[width=0.9\linewidth]{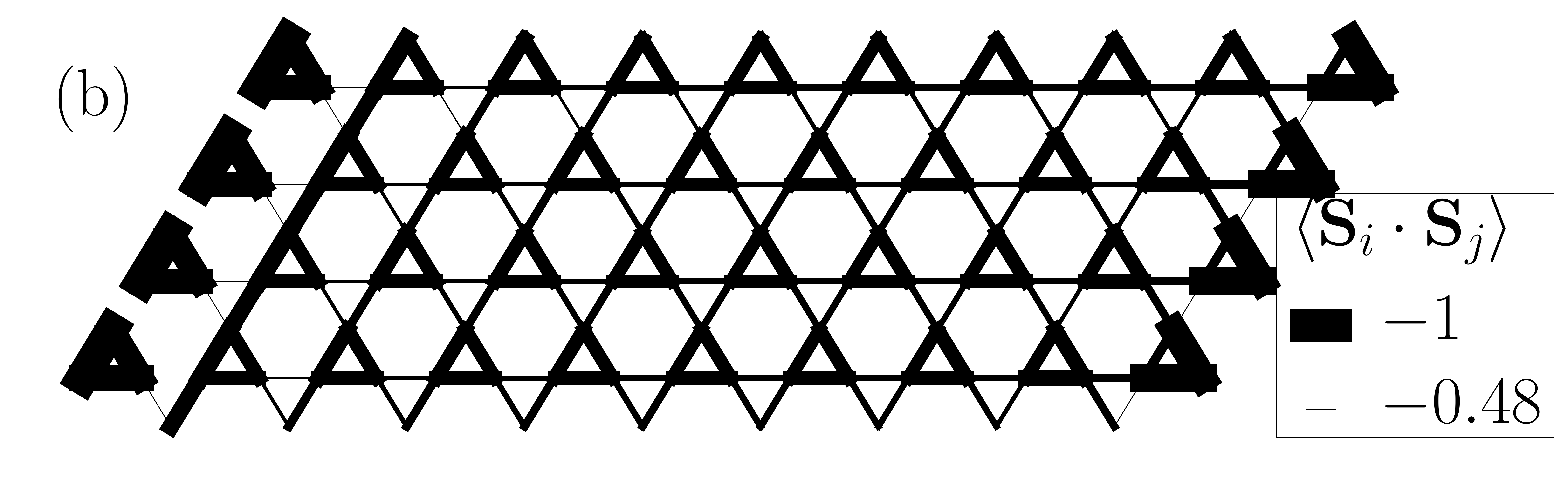}
\includegraphics[width=0.9\linewidth]{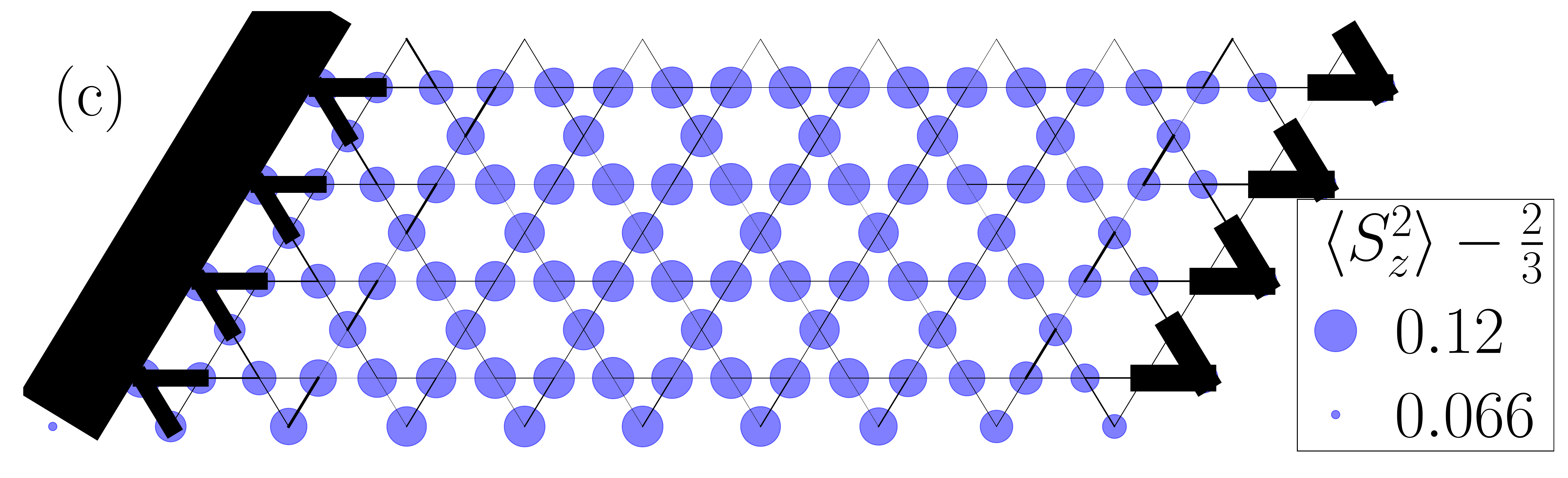}
\caption{(Color online) %\CMC{Add (a). Fixed according to the new figures. Hitesh please check if it is correct.}
(a) Energy spectrum of the 18b site cluster (organized by total $S_z$) from exact diagonalization, 
as a function of $J_R$ with $J_{bq}=-0.529 J_{R}$, fixing $J=9.2$ meV, the parameter set relevant for $U=3$ eV. 
The locations of the trimerized (T) and spin-nematic (SN) regions are indicated. 
The ground state in the HT structure corresponds to a spin-nematic (ferroquadrupolar) ground state with a $S=2$ excitation. 
(b,c) DMRG results for the trimerized and quadrupolar order parameters for the $S=1$ model with bilinear, 
biquadratic and ring-exchange terms with (b) $J_R/J=J_{bq}/J=0$ and (c) $J_R/J\approx-1.89J_{bq}/J\approx 0.37$ (the same as in Fig.~\ref{fig:fig2}) respectively.
The width of the bonds (radius of the circles) are proportional to $\langle \mathbf{S}_i \cdot \mathbf{S}_j \rangle$ ($\langle S_z^2\rangle-\frac{2}{3}$).
The reference values in the text boxes are valid for both cases.}
\label{fig:ED_DMRG} 
\end{figure}	

While the observation that $J$ is large and $J_{bq}$ and $J_R$ have opposite signs is robust, the values of $J_{bq}/J_R$ and $J_R/J$ ratios depend on the choice of $U$ we employ in the DFT+$U$ calculations \cite{supplement}. For this reason, it is necessary to perform the ED calculations for a range of parameter values. We scan the line in parameter space of the Hamiltonian in Eq.~\ref{eq:hamiltonian} with $J$ fixed to $9.2$ meV, varying $J_{R}$ with $J_{bq}=-0.529 J_{R}$. 
(Another scan for $J_R=-J_{bq}$, relevant for the $U=4$ eV parameters gives similar results \cite{supplement}.) 
Fig.~\ref{fig:ED_DMRG}a shows the energy spectrum (with the ground state energy subtracted for the corresponding choice of parameters) 
for the 18 site cluster as a function of $J_{R}$. There is a closing of energy scales, which signals the occurrence of a phase transition at 
$J_{R}\approx 1 $ meV. (This phase boundary is consistent with that we obtain for a 12 site kagome cluster (not shown), which suggests that the finite size effects are probably not important.) The lowest lying excitation in the large $J_R$ regime has $S=2$; consistent with the existence of a quadrupolar phase. Thus, the model with $J_R=-0.529J_{bq}$ is qualitatively similar to the model with $J_{R}=0$ and negative $J_{bq}$, and the magnetic ground state of \ntc in HT structure is quadrupolar. 

To build further confidence and confirm these assertions, we perform large scale DMRG calculations on XC8-3 cylinders with 
the open (periodic) boundaries along the long (short) direction. The open boundaries are chosen to match the trimer order. 
We explicitly measure the trimerized order parameter (defined as the difference of bond energies on up and down triangles i.e. 
$\sum_{\Delta} \mathbf{S}_i \cdot \mathbf{S}_j - \sum_{\nabla} \mathbf{S}_i \cdot \mathbf{S}_j$) and the quadrupolar order parameter $\langle S_z^2\rangle - \frac{2}{3}$. Results presented in Fig.~\ref{fig:ED_DMRG}b for $J_{R}/J=-J_{bq}/J=0$ confirm previous findings that the ground state is trimerized in the absence of biquadratic coupling. On the other hand, for $J_R/J\approx-1.89J_{bq}/J\approx 0.37$ (the same as in Fig.~\ref{fig:fig2}), a uniform non zero quadrupolar order parameter is observed throughout the bulk of our finite size sample, confirming the results obtained from ED (Fig.~\ref{fig:ED_DMRG}a). 

\prlsec{Emergence of trimerized phase and the role of lattice distortions}
Our discussion so far has focused mainly on the HT phase of the crystal structure. We now discuss what drives the instability towards the LT phase. The magnetic Hamiltonian of the HT phase by itself does not give rise to such an instability at low temperature, since both ED and DMRG calculations predict a nematic phase. 
It is possible that there is a lattice instability driven by not magnetism but rather by crystal chemistry, such as those in prototypical ferroelectrics like BaTiO$_3$ \cite{Cohen1992}. This type of a instability in \ntc would show up in DFT calculations as an unstable (imaginary frequency) phonon mode that transforms as $\Gamma_2^-$ irreducible representation (irrep) \cite{Kelly2019}. (While DFT is not capable of capturing the quantum magnetic phases at play here, it is expected to reproduce hybridization between atoms and other effects that gives rise to simple lattice instabilities.) Interestingly, our DFT calculations show no instability or soft mode unless an unphysical electronic structure is imposed \cite{supplement}. This suggests that there is no lattice instability towards trimerization either. 

What our calculations so far do not take into account is the spin--lattice coupling present in this material. Spin--lattice coupling is often considered in the context of materials with classical spin orders, where the changes in the crystal structure leads to differences in the magnetic energy through the dependence of exchange parameters to atomic positions. (See, for example, Refs.~\cite{Birol2012, Wysocki2016}.) In \ntc, the Wannier models discussed previously suggest that in the trimerized LT crystal structure, the relative strength of the biquadratic exchange to Heisenberg exchange, $J_{bq}/J$, is suppressed, and hence, the trimerized magnetic phase is favored more strongly in the LT phase compared to the HT phase. In other words, \textit{the spin--lattice coupling in \ntc favors the trimerized phase}, and we surmise that it is the driving force of the trimerization in this compound. 
The phase transition of \ntc driven by spin--lattice coupling can be considered as parallel to other compounds like VO$_2$, where a "chicken and egg" debate is still ongoing 
because the interactions between the correlated electronic states and the details of the lattice Hamiltonian give rise to concurrent electronic and crystal structural phase transitions \cite{Zylbersztejn1975,Wentzcovitch1994,Haverkort2005,Biermann2005,Weber2012,Gray2016,Quackenbush2013,Huffman2017,Vlad_VO2}. 

This instability, or any signature thereof, is not seen in DFT due to multiple inter-related reasons: 1) DFT, being a mean field theory, cannot capture the trimerized magnetic state of the electrons, and 2) The DFT+U implementations, developed to reproduce electrons localized on atoms, cannot capture the physics of electrons delocalized on trimers of Ti atoms. It is also highly likely that 3) since the transition is not second order, there may be no soft phonon mode responsible for it. We finally, note that in typical magnetic systems, spin Peierls distortions leads to an only $\sim$1-3\% change in the lattice constants, in contrast in 
\ntc the change is of the order of 10\%! 

An important feature of the $\Gamma_2^-$ structural distortion that connects the HT and LT structural phases is that it is polar \cite{Kelly2019}, and hence it couples to external electric fields in bilinear order. Since this lattice distortion is necessary for the trimerized phase to be more favorable than the quadrupolar phase, it might be possible to use electric fields at low temperatures to change the crystal structure enough to induce a transition to the quadrupolar phase. We surmise that this might be a viable strategy to probe a possible quantum critical point between these two magnetic phases.

\prlsec{Conclusions}
We have performed a theoretical and numerical analysis of the spin-1 Kagome compound \ntc using a combination of DFT, ED and DMRG calculations. 
We found that this compound has a complex magnetic Hamiltonian, which includes nearest neighbor biquadratic and ring exchange terms, in addition to strong antiferromagnetic Heisenberg interactions. ED and DMRG simulations agree on that depending on the strength of biquadratic and ring exchange terms, this Hamiltonian can give rise to either quadrupolar nematic or trimerized magnetic phases. We surmise that the magneto-structural transition observed in this compound is driven by the spin--lattice coupling, which favors the coexistence of the breathing distortion of the Kagome lattice and the trimerized magnetic phase. 

Our study underlines the importance of non-Heisenberg terms and lattice effects in the study of quantum magnetic materials, and shows that the spin--lattice coupling can lead to phase transitions that cannot be understood by studying magnetic or lattice Hamiltonians by themselves. This is similar to the well studied correlated compound, VO$_2$, which cannot be understood by studying the electronic or lattice subsystems alone. 

\begin{acknowledgments}
\prlsec{Acknowledgments} We thank Z. Kelly, T. McQueen, W. Ku, V. Dobrosavljevic, K. Yang, E. Manousakis, K. Plumb, C. Broholm, C. Hickey and Y. Kim for discussions. We also thank T. McQueen for introducing 
us to this material. HJC thanks A. Lauchli and (late) C.L. Henley for an earlier collaboration on the $S=1$ kagome system. HJC was supported by 
start up funds from Florida State University and the National High Magnetic Field Laboratory. We also thank the Research Computing Cluster (RCC) at Florida State University 
and XSEDE allocation (DMR190020) for computing resources. The National High Magnetic Field Laboratory is supported by the National Science Foundation through 
NSF/DMR-1644779 and the state of Florida. The DMRG calculations were performed using the ITensor C++ library (version 2.1.1)~\cite{ITensor}. The work at the University of Minnesota was supported by NSF DMREF Grant No. DMR-1629260. We acknowledge the Minnesota Supercomputing Institute for providing resources for the first principles calculations reported within this paper. 
%The DMRG calculations were performed using the ITensor C++ library (version 2.1.1), https://itensor.org/.
\end{acknowledgments}

\bibliography{refs}

\begin{thebibliography}{62}%
\makeatletter
\providecommand \@ifxundefined [1]{%
 \@ifx{#1\undefined}
}%
\providecommand \@ifnum [1]{%
 \ifnum #1\expandafter \@firstoftwo
 \else \expandafter \@secondoftwo
 \fi
}%
\providecommand \@ifx [1]{%
 \ifx #1\expandafter \@firstoftwo
 \else \expandafter \@secondoftwo
 \fi
}%
\providecommand \natexlab [1]{#1}%
\providecommand \enquote  [1]{``#1''}%
\providecommand \bibnamefont  [1]{#1}%
\providecommand \bibfnamefont [1]{#1}%
\providecommand \citenamefont [1]{#1}%
\providecommand \href@noop [0]{\@secondoftwo}%
\providecommand \href [0]{\begingroup \@sanitize@url \@href}%
\providecommand \@href[1]{\@@startlink{#1}\@@href}%
\providecommand \@@href[1]{\endgroup#1\@@endlink}%
\providecommand \@sanitize@url [0]{\catcode `\\12\catcode `\$12\catcode
  `\&12\catcode `\#12\catcode `\^12\catcode `\_12\catcode `\%12\relax}%
\providecommand \@@startlink[1]{}%
\providecommand \@@endlink[0]{}%
\providecommand \url  [0]{\begingroup\@sanitize@url \@url }%
\providecommand \@url [1]{\endgroup\@href {#1}{\urlprefix }}%
\providecommand \urlprefix  [0]{URL }%
\providecommand \Eprint [0]{\href }%
\providecommand \doibase [0]{http://dx.doi.org/}%
\providecommand \selectlanguage [0]{\@gobble}%
\providecommand \bibinfo  [0]{\@secondoftwo}%
\providecommand \bibfield  [0]{\@secondoftwo}%
\providecommand \translation [1]{[#1]}%
\providecommand \BibitemOpen [0]{}%
\providecommand \bibitemStop [0]{}%
\providecommand \bibitemNoStop [0]{.\EOS\space}%
\providecommand \EOS [0]{\spacefactor3000\relax}%
\providecommand \BibitemShut  [1]{\csname bibitem#1\endcsname}%
\let\auto@bib@innerbib\@empty
%</preamble>
\bibitem [{\citenamefont {Helton}\ \emph {et~al.}(2007)\citenamefont {Helton},
  \citenamefont {Matan}, \citenamefont {Shores}, \citenamefont {Nytko},
  \citenamefont {Bartlett}, \citenamefont {Yoshida}, \citenamefont {Takano},
  \citenamefont {Suslov}, \citenamefont {Qiu}, \citenamefont {Chung},
  \citenamefont {Nocera},\ and\ \citenamefont {Lee}}]{Nocera_Kagome_2007}%
  \BibitemOpen
  \bibfield  {author} {\bibinfo {author} {\bibfnamefont {J.~S.}\ \bibnamefont
  {Helton}}, \bibinfo {author} {\bibfnamefont {K.}~\bibnamefont {Matan}},
  \bibinfo {author} {\bibfnamefont {M.~P.}\ \bibnamefont {Shores}}, \bibinfo
  {author} {\bibfnamefont {E.~A.}\ \bibnamefont {Nytko}}, \bibinfo {author}
  {\bibfnamefont {B.~M.}\ \bibnamefont {Bartlett}}, \bibinfo {author}
  {\bibfnamefont {Y.}~\bibnamefont {Yoshida}}, \bibinfo {author} {\bibfnamefont
  {Y.}~\bibnamefont {Takano}}, \bibinfo {author} {\bibfnamefont
  {A.}~\bibnamefont {Suslov}}, \bibinfo {author} {\bibfnamefont
  {Y.}~\bibnamefont {Qiu}}, \bibinfo {author} {\bibfnamefont {J.-H.}\
  \bibnamefont {Chung}}, \bibinfo {author} {\bibfnamefont {D.~G.}\ \bibnamefont
  {Nocera}}, \ and\ \bibinfo {author} {\bibfnamefont {Y.~S.}\ \bibnamefont
  {Lee}},\ }\href {\doibase 10.1103/PhysRevLett.98.107204} {\bibfield
  {journal} {\bibinfo  {journal} {Phys. Rev. Lett.}\ }\textbf {\bibinfo
  {volume} {98}},\ \bibinfo {pages} {107204} (\bibinfo {year}
  {2007})}\BibitemShut {NoStop}%
\bibitem [{YS_()}]{YS_Lee_2012}%
  \BibitemOpen
  \href@noop {} {}\bibinfo {note} {Tian-Heng Han, Joel S. Helton, Shaoyan Chu,
  Daniel G. Nocera, Jose A. Rodriguez-Rivera, Collin Broholm and Young S. Lee,
  \emph{Nature} 492, 406-410 (2012)}\BibitemShut {NoStop}%
\bibitem [{\citenamefont {Zorko}\ \emph {et~al.}(2017)\citenamefont {Zorko},
  \citenamefont {Herak}, \citenamefont {Gomil\ifmmode~\check{s}\else
  \v{s}\fi{}ek}, \citenamefont {van Tol}, \citenamefont {Vel\'azquez},
  \citenamefont {Khuntia}, \citenamefont {Bert},\ and\ \citenamefont
  {Mendels}}]{Zorko_2017}%
  \BibitemOpen
  \bibfield  {author} {\bibinfo {author} {\bibfnamefont {A.}~\bibnamefont
  {Zorko}}, \bibinfo {author} {\bibfnamefont {M.}~\bibnamefont {Herak}},
  \bibinfo {author} {\bibfnamefont {M.}~\bibnamefont
  {Gomil\ifmmode~\check{s}\else \v{s}\fi{}ek}}, \bibinfo {author}
  {\bibfnamefont {J.}~\bibnamefont {van Tol}}, \bibinfo {author} {\bibfnamefont
  {M.}~\bibnamefont {Vel\'azquez}}, \bibinfo {author} {\bibfnamefont
  {P.}~\bibnamefont {Khuntia}}, \bibinfo {author} {\bibfnamefont
  {F.}~\bibnamefont {Bert}}, \ and\ \bibinfo {author} {\bibfnamefont
  {P.}~\bibnamefont {Mendels}},\ }\href {\doibase
  10.1103/PhysRevLett.118.017202} {\bibfield  {journal} {\bibinfo  {journal}
  {Phys. Rev. Lett.}\ }\textbf {\bibinfo {volume} {118}},\ \bibinfo {pages}
  {017202} (\bibinfo {year} {2017})}\BibitemShut {NoStop}%
\bibitem [{\citenamefont {Yan}\ \emph {et~al.}(2011)\citenamefont {Yan},
  \citenamefont {Huse},\ and\ \citenamefont {White}}]{White_kagome}%
  \BibitemOpen
  \bibfield  {author} {\bibinfo {author} {\bibfnamefont {S.}~\bibnamefont
  {Yan}}, \bibinfo {author} {\bibfnamefont {D.~A.}\ \bibnamefont {Huse}}, \
  and\ \bibinfo {author} {\bibfnamefont {S.~R.}\ \bibnamefont {White}},\ }\href
  {\doibase 10.1126/science.1201080} {\bibfield  {journal} {\bibinfo  {journal}
  {Science}\ }\textbf {\bibinfo {volume} {332}},\ \bibinfo {pages} {1173}
  (\bibinfo {year} {2011})},\ \Eprint
  {http://arxiv.org/abs/http://www.sciencemag.org/content/332/6034/1173.full.pdf}
  {http://www.sciencemag.org/content/332/6034/1173.full.pdf} \BibitemShut
  {NoStop}%
\bibitem [{\citenamefont {Ran}\ \emph {et~al.}(2007)\citenamefont {Ran},
  \citenamefont {Hermele}, \citenamefont {Lee},\ and\ \citenamefont
  {Wen}}]{Wen_Kagome}%
  \BibitemOpen
  \bibfield  {author} {\bibinfo {author} {\bibfnamefont {Y.}~\bibnamefont
  {Ran}}, \bibinfo {author} {\bibfnamefont {M.}~\bibnamefont {Hermele}},
  \bibinfo {author} {\bibfnamefont {P.~A.}\ \bibnamefont {Lee}}, \ and\
  \bibinfo {author} {\bibfnamefont {X.-G.}\ \bibnamefont {Wen}},\ }\href
  {\doibase 10.1103/PhysRevLett.98.117205} {\bibfield  {journal} {\bibinfo
  {journal} {Phys. Rev. Lett.}\ }\textbf {\bibinfo {volume} {98}},\ \bibinfo
  {pages} {117205} (\bibinfo {year} {2007})}\BibitemShut {NoStop}%
\bibitem [{\citenamefont {Depenbrock}\ \emph {et~al.}(2012)\citenamefont
  {Depenbrock}, \citenamefont {McCulloch},\ and\ \citenamefont
  {Schollw\"ock}}]{Depenbrock_Kagome}%
  \BibitemOpen
  \bibfield  {author} {\bibinfo {author} {\bibfnamefont {S.}~\bibnamefont
  {Depenbrock}}, \bibinfo {author} {\bibfnamefont {I.~P.}\ \bibnamefont
  {McCulloch}}, \ and\ \bibinfo {author} {\bibfnamefont {U.}~\bibnamefont
  {Schollw\"ock}},\ }\href {\doibase 10.1103/PhysRevLett.109.067201} {\bibfield
   {journal} {\bibinfo  {journal} {Phys. Rev. Lett.}\ }\textbf {\bibinfo
  {volume} {109}},\ \bibinfo {pages} {067201} (\bibinfo {year}
  {2012})}\BibitemShut {NoStop}%
\bibitem [{\citenamefont {Iqbal}\ \emph {et~al.}(2013)\citenamefont {Iqbal},
  \citenamefont {Becca}, \citenamefont {Sorella},\ and\ \citenamefont
  {Poilblanc}}]{Iqbal_kagome}%
  \BibitemOpen
  \bibfield  {author} {\bibinfo {author} {\bibfnamefont {Y.}~\bibnamefont
  {Iqbal}}, \bibinfo {author} {\bibfnamefont {F.}~\bibnamefont {Becca}},
  \bibinfo {author} {\bibfnamefont {S.}~\bibnamefont {Sorella}}, \ and\
  \bibinfo {author} {\bibfnamefont {D.}~\bibnamefont {Poilblanc}},\ }\href
  {\doibase 10.1103/PhysRevB.87.060405} {\bibfield  {journal} {\bibinfo
  {journal} {Phys. Rev. B}\ }\textbf {\bibinfo {volume} {87}},\ \bibinfo
  {pages} {060405} (\bibinfo {year} {2013})}\BibitemShut {NoStop}%
\bibitem [{\citenamefont {Norman}(2016)}]{Norman_2016}%
  \BibitemOpen
  \bibfield  {author} {\bibinfo {author} {\bibfnamefont {M.~R.}\ \bibnamefont
  {Norman}},\ }\href {\doibase 10.1103/RevModPhys.88.041002} {\bibfield
  {journal} {\bibinfo  {journal} {Rev. Mod. Phys.}\ }\textbf {\bibinfo {volume}
  {88}},\ \bibinfo {pages} {041002} (\bibinfo {year} {2016})}\BibitemShut
  {NoStop}%
\bibitem [{\citenamefont {Kumar}\ \emph {et~al.}(2016)\citenamefont {Kumar},
  \citenamefont {Changlani}, \citenamefont {Clark},\ and\ \citenamefont
  {Fradkin}}]{Kumar_PRB16}%
  \BibitemOpen
  \bibfield  {author} {\bibinfo {author} {\bibfnamefont {K.}~\bibnamefont
  {Kumar}}, \bibinfo {author} {\bibfnamefont {H.~J.}\ \bibnamefont
  {Changlani}}, \bibinfo {author} {\bibfnamefont {B.~K.}\ \bibnamefont
  {Clark}}, \ and\ \bibinfo {author} {\bibfnamefont {E.}~\bibnamefont
  {Fradkin}},\ }\href {\doibase 10.1103/PhysRevB.94.134410} {\bibfield
  {journal} {\bibinfo  {journal} {Phys. Rev. B}\ }\textbf {\bibinfo {volume}
  {94}},\ \bibinfo {pages} {134410} (\bibinfo {year} {2016})}\BibitemShut
  {NoStop}%
\bibitem [{\citenamefont {Changlani}\ \emph {et~al.}(2018)\citenamefont
  {Changlani}, \citenamefont {Kochkov}, \citenamefont {Kumar}, \citenamefont
  {Clark},\ and\ \citenamefont {Fradkin}}]{Changlani_PRL18}%
  \BibitemOpen
  \bibfield  {author} {\bibinfo {author} {\bibfnamefont {H.~J.}\ \bibnamefont
  {Changlani}}, \bibinfo {author} {\bibfnamefont {D.}~\bibnamefont {Kochkov}},
  \bibinfo {author} {\bibfnamefont {K.}~\bibnamefont {Kumar}}, \bibinfo
  {author} {\bibfnamefont {B.~K.}\ \bibnamefont {Clark}}, \ and\ \bibinfo
  {author} {\bibfnamefont {E.}~\bibnamefont {Fradkin}},\ }\href {\doibase
  10.1103/PhysRevLett.120.117202} {\bibfield  {journal} {\bibinfo  {journal}
  {Phys. Rev. Lett.}\ }\textbf {\bibinfo {volume} {120}},\ \bibinfo {pages}
  {117202} (\bibinfo {year} {2018})}\BibitemShut {NoStop}%
\bibitem [{\citenamefont {Hida}(2000)}]{Hida}%
  \BibitemOpen
  \bibfield  {author} {\bibinfo {author} {\bibfnamefont {K.}~\bibnamefont
  {Hida}},\ }\href {\doibase 10.1143/JPSJ.69.4003} {\bibfield  {journal}
  {\bibinfo  {journal} {Journal of the Physical Society of Japan}\ }\textbf
  {\bibinfo {volume} {69}},\ \bibinfo {pages} {4003} (\bibinfo {year}
  {2000})}\BibitemShut {NoStop}%
\bibitem [{\citenamefont {G\"otze}\ \emph {et~al.}(2011)\citenamefont
  {G\"otze}, \citenamefont {Farnell}, \citenamefont {Bishop}, \citenamefont
  {Li},\ and\ \citenamefont {Richter}}]{Gotze}%
  \BibitemOpen
  \bibfield  {author} {\bibinfo {author} {\bibfnamefont {O.}~\bibnamefont
  {G\"otze}}, \bibinfo {author} {\bibfnamefont {D.~J.~J.}\ \bibnamefont
  {Farnell}}, \bibinfo {author} {\bibfnamefont {R.~F.}\ \bibnamefont {Bishop}},
  \bibinfo {author} {\bibfnamefont {P.~H.~Y.}\ \bibnamefont {Li}}, \ and\
  \bibinfo {author} {\bibfnamefont {J.}~\bibnamefont {Richter}},\ }\href
  {\doibase 10.1103/PhysRevB.84.224428} {\bibfield  {journal} {\bibinfo
  {journal} {Phys. Rev. B}\ }\textbf {\bibinfo {volume} {84}},\ \bibinfo
  {pages} {224428} (\bibinfo {year} {2011})}\BibitemShut {NoStop}%
\bibitem [{\citenamefont {Arovas}(2008)}]{Arovas}%
  \BibitemOpen
  \bibfield  {author} {\bibinfo {author} {\bibfnamefont {D.~P.}\ \bibnamefont
  {Arovas}},\ }\href {\doibase 10.1103/PhysRevB.77.104404} {\bibfield
  {journal} {\bibinfo  {journal} {Phys. Rev. B}\ }\textbf {\bibinfo {volume}
  {77}},\ \bibinfo {pages} {104404} (\bibinfo {year} {2008})}\BibitemShut
  {NoStop}%
\bibitem [{\citenamefont {Corboz}\ \emph {et~al.}(2012)\citenamefont {Corboz},
  \citenamefont {Penc}, \citenamefont {Mila},\ and\ \citenamefont
  {L\"auchli}}]{Corboz}%
  \BibitemOpen
  \bibfield  {author} {\bibinfo {author} {\bibfnamefont {P.}~\bibnamefont
  {Corboz}}, \bibinfo {author} {\bibfnamefont {K.}~\bibnamefont {Penc}},
  \bibinfo {author} {\bibfnamefont {F.}~\bibnamefont {Mila}}, \ and\ \bibinfo
  {author} {\bibfnamefont {A.~M.}\ \bibnamefont {L\"auchli}},\ }\href {\doibase
  10.1103/PhysRevB.86.041106} {\bibfield  {journal} {\bibinfo  {journal} {Phys.
  Rev. B}\ }\textbf {\bibinfo {volume} {86}},\ \bibinfo {pages} {041106}
  (\bibinfo {year} {2012})}\BibitemShut {NoStop}%
\bibitem [{\citenamefont {Changlani}\ and\ \citenamefont
  {L\"auchli}(2015)}]{Changlani_PRB15}%
  \BibitemOpen
  \bibfield  {author} {\bibinfo {author} {\bibfnamefont {H.~J.}\ \bibnamefont
  {Changlani}}\ and\ \bibinfo {author} {\bibfnamefont {A.~M.}\ \bibnamefont
  {L\"auchli}},\ }\href {\doibase 10.1103/PhysRevB.91.100407} {\bibfield
  {journal} {\bibinfo  {journal} {Phys. Rev. B}\ }\textbf {\bibinfo {volume}
  {91}},\ \bibinfo {pages} {100407} (\bibinfo {year} {2015})}\BibitemShut
  {NoStop}%
\bibitem [{\citenamefont {Liu}\ \emph {et~al.}(2015)\citenamefont {Liu},
  \citenamefont {Li}, \citenamefont {Weichselbaum}, \citenamefont {von Delft},\
  and\ \citenamefont {Su}}]{vondelft_PRB15}%
  \BibitemOpen
  \bibfield  {author} {\bibinfo {author} {\bibfnamefont {T.}~\bibnamefont
  {Liu}}, \bibinfo {author} {\bibfnamefont {W.}~\bibnamefont {Li}}, \bibinfo
  {author} {\bibfnamefont {A.}~\bibnamefont {Weichselbaum}}, \bibinfo {author}
  {\bibfnamefont {J.}~\bibnamefont {von Delft}}, \ and\ \bibinfo {author}
  {\bibfnamefont {G.}~\bibnamefont {Su}},\ }\href {\doibase
  10.1103/PhysRevB.91.060403} {\bibfield  {journal} {\bibinfo  {journal} {Phys.
  Rev. B}\ }\textbf {\bibinfo {volume} {91}},\ \bibinfo {pages} {060403}
  (\bibinfo {year} {2015})}\BibitemShut {NoStop}%
\bibitem [{\citenamefont {Ghosh}\ \emph {et~al.}(2016)\citenamefont {Ghosh},
  \citenamefont {Verma},\ and\ \citenamefont {Kumar}}]{Ghosh_PRB16}%
  \BibitemOpen
  \bibfield  {author} {\bibinfo {author} {\bibfnamefont {P.}~\bibnamefont
  {Ghosh}}, \bibinfo {author} {\bibfnamefont {A.~K.}\ \bibnamefont {Verma}}, \
  and\ \bibinfo {author} {\bibfnamefont {B.}~\bibnamefont {Kumar}},\ }\href
  {\doibase 10.1103/PhysRevB.93.014427} {\bibfield  {journal} {\bibinfo
  {journal} {Phys. Rev. B}\ }\textbf {\bibinfo {volume} {93}},\ \bibinfo
  {pages} {014427} (\bibinfo {year} {2016})}\BibitemShut {NoStop}%
\bibitem [{\citenamefont {Plumb}\ \emph {et~al.}(2019)\citenamefont {Plumb},
  \citenamefont {Changlani}, \citenamefont {Scheie}, \citenamefont {Zhang},
  \citenamefont {Krizan}, \citenamefont {Rodriguez-Rivera}, \citenamefont
  {Qiu}, \citenamefont {Winn}, \citenamefont {Cava},\ and\ \citenamefont
  {Broholm}}]{Plumb_NatPhys19}%
  \BibitemOpen
  \bibfield  {author} {\bibinfo {author} {\bibfnamefont {K.~W.}\ \bibnamefont
  {Plumb}}, \bibinfo {author} {\bibfnamefont {H.~J.}\ \bibnamefont
  {Changlani}}, \bibinfo {author} {\bibfnamefont {A.}~\bibnamefont {Scheie}},
  \bibinfo {author} {\bibfnamefont {S.}~\bibnamefont {Zhang}}, \bibinfo
  {author} {\bibfnamefont {J.~W.}\ \bibnamefont {Krizan}}, \bibinfo {author}
  {\bibfnamefont {J.~A.}\ \bibnamefont {Rodriguez-Rivera}}, \bibinfo {author}
  {\bibfnamefont {Y.}~\bibnamefont {Qiu}}, \bibinfo {author} {\bibfnamefont
  {B.}~\bibnamefont {Winn}}, \bibinfo {author} {\bibfnamefont {R.~J.}\
  \bibnamefont {Cava}}, \ and\ \bibinfo {author} {\bibfnamefont {C.~L.}\
  \bibnamefont {Broholm}},\ }\href {\doibase 10.1038/s41567-018-0317-3}
  {\bibfield  {journal} {\bibinfo  {journal} {Nature Physics}\ }\textbf
  {\bibinfo {volume} {15}},\ \bibinfo {pages} {54} (\bibinfo {year}
  {2019})}\BibitemShut {NoStop}%
\bibitem [{\citenamefont {Zhang}\ \emph {et~al.}(2019)\citenamefont {Zhang},
  \citenamefont {Changlani}, \citenamefont {Plumb}, \citenamefont
  {Tchernyshyov},\ and\ \citenamefont {Moessner}}]{Zhang_PRL19}%
  \BibitemOpen
  \bibfield  {author} {\bibinfo {author} {\bibfnamefont {S.}~\bibnamefont
  {Zhang}}, \bibinfo {author} {\bibfnamefont {H.~J.}\ \bibnamefont
  {Changlani}}, \bibinfo {author} {\bibfnamefont {K.~W.}\ \bibnamefont
  {Plumb}}, \bibinfo {author} {\bibfnamefont {O.}~\bibnamefont {Tchernyshyov}},
  \ and\ \bibinfo {author} {\bibfnamefont {R.}~\bibnamefont {Moessner}},\
  }\href {\doibase 10.1103/PhysRevLett.122.167203} {\bibfield  {journal}
  {\bibinfo  {journal} {Phys. Rev. Lett.}\ }\textbf {\bibinfo {volume} {122}},\
  \bibinfo {pages} {167203} (\bibinfo {year} {2019})}\BibitemShut {NoStop}%
\bibitem [{\citenamefont {Iqbal}\ \emph {et~al.}(2019)\citenamefont {Iqbal},
  \citenamefont {M\"uller}, \citenamefont {Ghosh}, \citenamefont {Gingras},
  \citenamefont {Jeschke}, \citenamefont {Rachel}, \citenamefont {Reuther},\
  and\ \citenamefont {Thomale}}]{Iqbal_PRX19}%
  \BibitemOpen
  \bibfield  {author} {\bibinfo {author} {\bibfnamefont {Y.}~\bibnamefont
  {Iqbal}}, \bibinfo {author} {\bibfnamefont {T.}~\bibnamefont {M\"uller}},
  \bibinfo {author} {\bibfnamefont {P.}~\bibnamefont {Ghosh}}, \bibinfo
  {author} {\bibfnamefont {M.~J.~P.}\ \bibnamefont {Gingras}}, \bibinfo
  {author} {\bibfnamefont {H.~O.}\ \bibnamefont {Jeschke}}, \bibinfo {author}
  {\bibfnamefont {S.}~\bibnamefont {Rachel}}, \bibinfo {author} {\bibfnamefont
  {J.}~\bibnamefont {Reuther}}, \ and\ \bibinfo {author} {\bibfnamefont
  {R.}~\bibnamefont {Thomale}},\ }\href {\doibase 10.1103/PhysRevX.9.011005}
  {\bibfield  {journal} {\bibinfo  {journal} {Phys. Rev. X}\ }\textbf {\bibinfo
  {volume} {9}},\ \bibinfo {pages} {011005} (\bibinfo {year}
  {2019})}\BibitemShut {NoStop}%
\bibitem [{\citenamefont {Nakatsuji}\ \emph {et~al.}(2005)\citenamefont
  {Nakatsuji}, \citenamefont {Nambu}, \citenamefont {Tonomura}, \citenamefont
  {Sakai}, \citenamefont {Jonas}, \citenamefont {Broholm}, \citenamefont
  {Tsunetsugu}, \citenamefont {Qiu},\ and\ \citenamefont
  {Maeno}}]{Nakatsuji_2005}%
  \BibitemOpen
  \bibfield  {author} {\bibinfo {author} {\bibfnamefont {S.}~\bibnamefont
  {Nakatsuji}}, \bibinfo {author} {\bibfnamefont {Y.}~\bibnamefont {Nambu}},
  \bibinfo {author} {\bibfnamefont {H.}~\bibnamefont {Tonomura}}, \bibinfo
  {author} {\bibfnamefont {O.}~\bibnamefont {Sakai}}, \bibinfo {author}
  {\bibfnamefont {S.}~\bibnamefont {Jonas}}, \bibinfo {author} {\bibfnamefont
  {C.}~\bibnamefont {Broholm}}, \bibinfo {author} {\bibfnamefont
  {H.}~\bibnamefont {Tsunetsugu}}, \bibinfo {author} {\bibfnamefont
  {Y.}~\bibnamefont {Qiu}}, \ and\ \bibinfo {author} {\bibfnamefont
  {Y.}~\bibnamefont {Maeno}},\ }\href {\doibase 10.1126/science.1114727}
  {\bibfield  {journal} {\bibinfo  {journal} {Science}\ }\textbf {\bibinfo
  {volume} {309}},\ \bibinfo {pages} {1697} (\bibinfo {year} {2005})},\ \Eprint
  {http://arxiv.org/abs/https://science.sciencemag.org/content/309/5741/1697.full.pdf}
  {https://science.sciencemag.org/content/309/5741/1697.full.pdf} \BibitemShut
  {NoStop}%
\bibitem [{\citenamefont {Cheng}\ \emph {et~al.}(2011)\citenamefont {Cheng},
  \citenamefont {Li}, \citenamefont {Balicas}, \citenamefont {Zhou},
  \citenamefont {Goodenough}, \citenamefont {Xu},\ and\ \citenamefont
  {Zhou}}]{Balicas_2011}%
  \BibitemOpen
  \bibfield  {author} {\bibinfo {author} {\bibfnamefont {J.~G.}\ \bibnamefont
  {Cheng}}, \bibinfo {author} {\bibfnamefont {G.}~\bibnamefont {Li}}, \bibinfo
  {author} {\bibfnamefont {L.}~\bibnamefont {Balicas}}, \bibinfo {author}
  {\bibfnamefont {J.~S.}\ \bibnamefont {Zhou}}, \bibinfo {author}
  {\bibfnamefont {J.~B.}\ \bibnamefont {Goodenough}}, \bibinfo {author}
  {\bibfnamefont {C.}~\bibnamefont {Xu}}, \ and\ \bibinfo {author}
  {\bibfnamefont {H.~D.}\ \bibnamefont {Zhou}},\ }\href {\doibase
  10.1103/PhysRevLett.107.197204} {\bibfield  {journal} {\bibinfo  {journal}
  {Phys. Rev. Lett.}\ }\textbf {\bibinfo {volume} {107}},\ \bibinfo {pages}
  {197204} (\bibinfo {year} {2011})}\BibitemShut {NoStop}%
\bibitem [{\citenamefont {Serbyn}\ \emph {et~al.}(2011)\citenamefont {Serbyn},
  \citenamefont {Senthil},\ and\ \citenamefont {Lee}}]{Serbyn_2011}%
  \BibitemOpen
  \bibfield  {author} {\bibinfo {author} {\bibfnamefont {M.}~\bibnamefont
  {Serbyn}}, \bibinfo {author} {\bibfnamefont {T.}~\bibnamefont {Senthil}}, \
  and\ \bibinfo {author} {\bibfnamefont {P.~A.}\ \bibnamefont {Lee}},\ }\href
  {\doibase 10.1103/PhysRevB.84.180403} {\bibfield  {journal} {\bibinfo
  {journal} {Phys. Rev. B}\ }\textbf {\bibinfo {volume} {84}},\ \bibinfo
  {pages} {180403} (\bibinfo {year} {2011})}\BibitemShut {NoStop}%
\bibitem [{\citenamefont {L\"auchli}\ \emph {et~al.}(2006)\citenamefont
  {L\"auchli}, \citenamefont {Mila},\ and\ \citenamefont
  {Penc}}]{Lauchli_Mila_Penc}%
  \BibitemOpen
  \bibfield  {author} {\bibinfo {author} {\bibfnamefont {A.}~\bibnamefont
  {L\"auchli}}, \bibinfo {author} {\bibfnamefont {F.}~\bibnamefont {Mila}}, \
  and\ \bibinfo {author} {\bibfnamefont {K.}~\bibnamefont {Penc}},\ }\href
  {\doibase 10.1103/PhysRevLett.97.087205} {\bibfield  {journal} {\bibinfo
  {journal} {Phys. Rev. Lett.}\ }\textbf {\bibinfo {volume} {97}},\ \bibinfo
  {pages} {087205} (\bibinfo {year} {2006})}\BibitemShut {NoStop}%
\bibitem [{\citenamefont {Kumar}\ \emph {et~al.}(2019)\citenamefont {Kumar},
  \citenamefont {Dey}, \citenamefont {Ette}, \citenamefont {Ramesha},
  \citenamefont {Chakraborty}, \citenamefont {Dasgupta}, \citenamefont {Orain},
  \citenamefont {Baines}, \citenamefont {T\'oth}, \citenamefont {Shahee},
  \citenamefont {Kundu}, \citenamefont {Prinz-Zwick}, \citenamefont {Gippius},
  \citenamefont {B\"uttgen}, \citenamefont {Gegenwart},\ and\ \citenamefont
  {Mahajan}}]{Mahajan_honeycomb}%
  \BibitemOpen
  \bibfield  {author} {\bibinfo {author} {\bibfnamefont {R.}~\bibnamefont
  {Kumar}}, \bibinfo {author} {\bibfnamefont {T.}~\bibnamefont {Dey}}, \bibinfo
  {author} {\bibfnamefont {P.~M.}\ \bibnamefont {Ette}}, \bibinfo {author}
  {\bibfnamefont {K.}~\bibnamefont {Ramesha}}, \bibinfo {author} {\bibfnamefont
  {A.}~\bibnamefont {Chakraborty}}, \bibinfo {author} {\bibfnamefont
  {I.}~\bibnamefont {Dasgupta}}, \bibinfo {author} {\bibfnamefont {J.~C.}\
  \bibnamefont {Orain}}, \bibinfo {author} {\bibfnamefont {C.}~\bibnamefont
  {Baines}}, \bibinfo {author} {\bibfnamefont {S.}~\bibnamefont {T\'oth}},
  \bibinfo {author} {\bibfnamefont {A.}~\bibnamefont {Shahee}}, \bibinfo
  {author} {\bibfnamefont {S.}~\bibnamefont {Kundu}}, \bibinfo {author}
  {\bibfnamefont {M.}~\bibnamefont {Prinz-Zwick}}, \bibinfo {author}
  {\bibfnamefont {A.~A.}\ \bibnamefont {Gippius}}, \bibinfo {author}
  {\bibfnamefont {N.}~\bibnamefont {B\"uttgen}}, \bibinfo {author}
  {\bibfnamefont {P.}~\bibnamefont {Gegenwart}}, \ and\ \bibinfo {author}
  {\bibfnamefont {A.~V.}\ \bibnamefont {Mahajan}},\ }\href {\doibase
  10.1103/PhysRevB.99.054417} {\bibfield  {journal} {\bibinfo  {journal} {Phys.
  Rev. B}\ }\textbf {\bibinfo {volume} {99}},\ \bibinfo {pages} {054417}
  (\bibinfo {year} {2019})}\BibitemShut {NoStop}%
\bibitem [{\citenamefont {Silverstein}\ \emph {et~al.}(2018)\citenamefont
  {Silverstein}, \citenamefont {Sinclair}, \citenamefont {Sharma},
  \citenamefont {Qiu}, \citenamefont {Heinmaa}, \citenamefont {Leitm\"ae},
  \citenamefont {Wiebe}, \citenamefont {Stern},\ and\ \citenamefont
  {Zhou}}]{Zhou_PRM18}%
  \BibitemOpen
  \bibfield  {author} {\bibinfo {author} {\bibfnamefont {H.~J.}\ \bibnamefont
  {Silverstein}}, \bibinfo {author} {\bibfnamefont {R.}~\bibnamefont
  {Sinclair}}, \bibinfo {author} {\bibfnamefont {A.}~\bibnamefont {Sharma}},
  \bibinfo {author} {\bibfnamefont {Y.}~\bibnamefont {Qiu}}, \bibinfo {author}
  {\bibfnamefont {I.}~\bibnamefont {Heinmaa}}, \bibinfo {author} {\bibfnamefont
  {A.}~\bibnamefont {Leitm\"ae}}, \bibinfo {author} {\bibfnamefont {C.~R.}\
  \bibnamefont {Wiebe}}, \bibinfo {author} {\bibfnamefont {R.}~\bibnamefont
  {Stern}}, \ and\ \bibinfo {author} {\bibfnamefont {H.}~\bibnamefont {Zhou}},\
  }\href {\doibase 10.1103/PhysRevMaterials.2.044006} {\bibfield  {journal}
  {\bibinfo  {journal} {Phys. Rev. Materials}\ }\textbf {\bibinfo {volume}
  {2}},\ \bibinfo {pages} {044006} (\bibinfo {year} {2018})}\BibitemShut
  {NoStop}%
\bibitem [{\citenamefont {Takagi}\ \emph {et~al.}(2017)\citenamefont {Takagi},
  \citenamefont {Aoyama}, \citenamefont {Hara}, \citenamefont {Sato},
  \citenamefont {Kimura},\ and\ \citenamefont {Wakabayashi}}]{Takagi_PRB17}%
  \BibitemOpen
  \bibfield  {author} {\bibinfo {author} {\bibfnamefont {E.}~\bibnamefont
  {Takagi}}, \bibinfo {author} {\bibfnamefont {T.}~\bibnamefont {Aoyama}},
  \bibinfo {author} {\bibfnamefont {S.}~\bibnamefont {Hara}}, \bibinfo {author}
  {\bibfnamefont {H.}~\bibnamefont {Sato}}, \bibinfo {author} {\bibfnamefont
  {T.}~\bibnamefont {Kimura}}, \ and\ \bibinfo {author} {\bibfnamefont
  {Y.}~\bibnamefont {Wakabayashi}},\ }\href {\doibase
  10.1103/PhysRevB.95.104416} {\bibfield  {journal} {\bibinfo  {journal} {Phys.
  Rev. B}\ }\textbf {\bibinfo {volume} {95}},\ \bibinfo {pages} {104416}
  (\bibinfo {year} {2017})}\BibitemShut {NoStop}%
\bibitem [{\citenamefont {Hinz}\ \emph {et~al.}(1995)\citenamefont {Hinz},
  \citenamefont {Meyer}, \citenamefont {Dedecke},\ and\ \citenamefont
  {Urland}}]{Hinz1995}%
  \BibitemOpen
  \bibfield  {author} {\bibinfo {author} {\bibfnamefont {D.~J.}\ \bibnamefont
  {Hinz}}, \bibinfo {author} {\bibfnamefont {G.}~\bibnamefont {Meyer}},
  \bibinfo {author} {\bibfnamefont {T.}~\bibnamefont {Dedecke}}, \ and\
  \bibinfo {author} {\bibfnamefont {W.}~\bibnamefont {Urland}},\ }\href
  {\doibase 10.1002/anie.199500711} {\bibfield  {journal} {\bibinfo  {journal}
  {Angewandte Chemie International Edition in English}\ }\textbf {\bibinfo
  {volume} {34}},\ \bibinfo {pages} {71} (\bibinfo {year} {1995})}\BibitemShut
  {NoStop}%
\bibitem [{\citenamefont {Hanni}\ \emph {et~al.}(2017)\citenamefont {Hanni},
  \citenamefont {Frontzek}, \citenamefont {Hauser}, \citenamefont
  {Cheptiakov},\ and\ \citenamefont {Kramer}}]{Hanni_17}%
  \BibitemOpen
  \bibfield  {author} {\bibinfo {author} {\bibfnamefont {N.}~\bibnamefont
  {Hanni}}, \bibinfo {author} {\bibfnamefont {M.~D.}\ \bibnamefont {Frontzek}},
  \bibinfo {author} {\bibfnamefont {J.}~\bibnamefont {Hauser}}, \bibinfo
  {author} {\bibfnamefont {D.}~\bibnamefont {Cheptiakov}}, \ and\ \bibinfo
  {author} {\bibfnamefont {K.}~\bibnamefont {Kramer}},\ }\href {\doibase
  10.1002/zaac.201700331} {\bibfield  {journal} {\bibinfo  {journal}
  {Zeitschrift fuer Anorganische und Allgemeine Chemie}\ }\textbf {\bibinfo
  {volume} {643}} (\bibinfo {year} {2017}),\
  10.1002/zaac.201700331}\BibitemShut {NoStop}%
\bibitem [{\citenamefont {Kelly}\ \emph {et~al.}(2019)\citenamefont {Kelly},
  \citenamefont {Tran},\ and\ \citenamefont {McQueen}}]{Kelly2019}%
  \BibitemOpen
  \bibfield  {author} {\bibinfo {author} {\bibfnamefont {Z.~A.}\ \bibnamefont
  {Kelly}}, \bibinfo {author} {\bibfnamefont {T.~T.}\ \bibnamefont {Tran}}, \
  and\ \bibinfo {author} {\bibfnamefont {T.~M.}\ \bibnamefont {McQueen}},\
  }\href {\doibase 10.1021/acs.inorgchem.9b01110} {\bibfield  {journal}
  {\bibinfo  {journal} {Inorganic Chemistry}\ ,\ \bibinfo {pages}
  {acs.inorgchem.9b01110}} (\bibinfo {year} {2019})}\BibitemShut {NoStop}%
\bibitem [{\citenamefont {Changlani}\ \emph {et~al.}(2015)\citenamefont
  {Changlani}, \citenamefont {Zheng},\ and\ \citenamefont
  {Wagner}}]{Changlani_JCP2015}%
  \BibitemOpen
  \bibfield  {author} {\bibinfo {author} {\bibfnamefont {H.~J.}\ \bibnamefont
  {Changlani}}, \bibinfo {author} {\bibfnamefont {H.}~\bibnamefont {Zheng}}, \
  and\ \bibinfo {author} {\bibfnamefont {L.~K.}\ \bibnamefont {Wagner}},\
  }\href {\doibase 10.1063/1.4927664} {\bibfield  {journal} {\bibinfo
  {journal} {The Journal of Chemical Physics}\ }\textbf {\bibinfo {volume}
  {143}},\ \bibinfo {pages} {102814} (\bibinfo {year} {2015})},\ \Eprint
  {http://arxiv.org/abs/https://doi.org/10.1063/1.4927664}
  {https://doi.org/10.1063/1.4927664} \BibitemShut {NoStop}%
\bibitem [{\citenamefont {Zheng}\ \emph {et~al.}(2018)\citenamefont {Zheng},
  \citenamefont {Changlani}, \citenamefont {Williams}, \citenamefont
  {Busemeyer},\ and\ \citenamefont {Wagner}}]{Zheng_FrontPhys2018}%
  \BibitemOpen
  \bibfield  {author} {\bibinfo {author} {\bibfnamefont {H.}~\bibnamefont
  {Zheng}}, \bibinfo {author} {\bibfnamefont {H.~J.}\ \bibnamefont
  {Changlani}}, \bibinfo {author} {\bibfnamefont {K.~T.}\ \bibnamefont
  {Williams}}, \bibinfo {author} {\bibfnamefont {B.}~\bibnamefont {Busemeyer}},
  \ and\ \bibinfo {author} {\bibfnamefont {L.~K.}\ \bibnamefont {Wagner}},\
  }\href {\doibase 10.3389/fphy.2018.00043} {\bibfield  {journal} {\bibinfo
  {journal} {Frontiers in Physics}\ }\textbf {\bibinfo {volume} {6}},\ \bibinfo
  {pages} {43} (\bibinfo {year} {2018})}\BibitemShut {NoStop}%
\bibitem [{\citenamefont {Fedorova}\ \emph {et~al.}(2015)\citenamefont
  {Fedorova}, \citenamefont {Ederer}, \citenamefont {Spaldin},\ and\
  \citenamefont {Scaramucci}}]{Spaldin}%
  \BibitemOpen
  \bibfield  {author} {\bibinfo {author} {\bibfnamefont {N.~S.}\ \bibnamefont
  {Fedorova}}, \bibinfo {author} {\bibfnamefont {C.}~\bibnamefont {Ederer}},
  \bibinfo {author} {\bibfnamefont {N.~A.}\ \bibnamefont {Spaldin}}, \ and\
  \bibinfo {author} {\bibfnamefont {A.}~\bibnamefont {Scaramucci}},\ }\href
  {\doibase 10.1103/PhysRevB.91.165122} {\bibfield  {journal} {\bibinfo
  {journal} {Phys. Rev. B}\ }\textbf {\bibinfo {volume} {91}},\ \bibinfo
  {pages} {165122} (\bibinfo {year} {2015})}\BibitemShut {NoStop}%
\bibitem [{\citenamefont {Jeschke}\ \emph {et~al.}(2013)\citenamefont
  {Jeschke}, \citenamefont {Salvat-Pujol},\ and\ \citenamefont
  {Valent\'{\i}}}]{Jeschke_Valenti}%
  \BibitemOpen
  \bibfield  {author} {\bibinfo {author} {\bibfnamefont {H.~O.}\ \bibnamefont
  {Jeschke}}, \bibinfo {author} {\bibfnamefont {F.}~\bibnamefont
  {Salvat-Pujol}}, \ and\ \bibinfo {author} {\bibfnamefont {R.}~\bibnamefont
  {Valent\'{\i}}},\ }\href {\doibase 10.1103/PhysRevB.88.075106} {\bibfield
  {journal} {\bibinfo  {journal} {Phys. Rev. B}\ }\textbf {\bibinfo {volume}
  {88}},\ \bibinfo {pages} {075106} (\bibinfo {year} {2013})}\BibitemShut
  {NoStop}%
\bibitem [{\citenamefont {Birol}\ \emph {et~al.}(2018)\citenamefont {Birol},
  \citenamefont {Haule},\ and\ \citenamefont {Vanderbilt}}]{Birol2018}%
  \BibitemOpen
  \bibfield  {author} {\bibinfo {author} {\bibfnamefont {T.}~\bibnamefont
  {Birol}}, \bibinfo {author} {\bibfnamefont {K.}~\bibnamefont {Haule}}, \ and\
  \bibinfo {author} {\bibfnamefont {D.}~\bibnamefont {Vanderbilt}},\ }\href
  {\doibase 10.1103/PhysRevB.98.134432} {\bibfield  {journal} {\bibinfo
  {journal} {Physical Review B}\ }\textbf {\bibinfo {volume} {98}},\ \bibinfo
  {pages} {134432} (\bibinfo {year} {2018})}\BibitemShut {NoStop}%
\bibitem [{\citenamefont {supplemental information for~further
  details.}()}]{supplement}%
  \BibitemOpen
  \bibfield  {author} {\bibinfo {author} {\bibfnamefont {S.}~\bibnamefont
  {supplemental information for~further details.}},\ }\href@noop {} {}\bibinfo
  {type} {Tech. Rep.}\BibitemShut {Stop}%
\bibitem [{\citenamefont {Kresse}\ and\ \citenamefont
  {Furthmuller}(1996)}]{VASP1}%
  \BibitemOpen
  \bibfield  {author} {\bibinfo {author} {\bibfnamefont {G.}~\bibnamefont
  {Kresse}}\ and\ \bibinfo {author} {\bibfnamefont {J.}~\bibnamefont
  {Furthmuller}},\ }\href@noop {} {\bibfield  {journal} {\bibinfo  {journal}
  {Computational Materials Science}\ }\textbf {\bibinfo {volume} {6}},\
  \bibinfo {pages} {15} (\bibinfo {year} {1996})}\BibitemShut {NoStop}%
\bibitem [{\citenamefont {Kresse}\ and\ \citenamefont
  {Furthm{\"{u}}ller}(1996)}]{VASP2}%
  \BibitemOpen
  \bibfield  {author} {\bibinfo {author} {\bibfnamefont {G.}~\bibnamefont
  {Kresse}}\ and\ \bibinfo {author} {\bibfnamefont {J.}~\bibnamefont
  {Furthm{\"{u}}ller}},\ }\href {\doibase 10.1103/PhysRevB.54.11169} {\bibfield
   {journal} {\bibinfo  {journal} {Physical Review B}\ }\textbf {\bibinfo
  {volume} {54}},\ \bibinfo {pages} {11169} (\bibinfo {year}
  {1996})}\BibitemShut {NoStop}%
\bibitem [{\citenamefont {Perdew}\ \emph {et~al.}(2008)\citenamefont {Perdew},
  \citenamefont {Ruzsinszky}, \citenamefont {Csonka}, \citenamefont {Vydrov},
  \citenamefont {Scuseria}, \citenamefont {Constantin}, \citenamefont {Zhou},\
  and\ \citenamefont {Burke}}]{PBEsol}%
  \BibitemOpen
  \bibfield  {author} {\bibinfo {author} {\bibfnamefont {J.~P.}\ \bibnamefont
  {Perdew}}, \bibinfo {author} {\bibfnamefont {A.}~\bibnamefont {Ruzsinszky}},
  \bibinfo {author} {\bibfnamefont {G.~I.}\ \bibnamefont {Csonka}}, \bibinfo
  {author} {\bibfnamefont {O.~A.}\ \bibnamefont {Vydrov}}, \bibinfo {author}
  {\bibfnamefont {G.~E.}\ \bibnamefont {Scuseria}}, \bibinfo {author}
  {\bibfnamefont {L.~A.}\ \bibnamefont {Constantin}}, \bibinfo {author}
  {\bibfnamefont {X.}~\bibnamefont {Zhou}}, \ and\ \bibinfo {author}
  {\bibfnamefont {K.}~\bibnamefont {Burke}},\ }\href {\doibase
  10.1103/PhysRevLett.100.136406} {\bibfield  {journal} {\bibinfo  {journal}
  {Physical Review Letters}\ }\textbf {\bibinfo {volume} {100}},\ \bibinfo
  {pages} {136406} (\bibinfo {year} {2008})}\BibitemShut {NoStop}%
\bibitem [{\citenamefont {Dudarev}\ \emph {et~al.}(1998)\citenamefont
  {Dudarev}, \citenamefont {Botton}, \citenamefont {Savrasov}, \citenamefont
  {Humphreys},\ and\ \citenamefont {Sutton}}]{Dudarev1998}%
  \BibitemOpen
  \bibfield  {author} {\bibinfo {author} {\bibfnamefont {S.~L.}\ \bibnamefont
  {Dudarev}}, \bibinfo {author} {\bibfnamefont {G.~A.}\ \bibnamefont {Botton}},
  \bibinfo {author} {\bibfnamefont {S.~Y.}\ \bibnamefont {Savrasov}}, \bibinfo
  {author} {\bibfnamefont {C.~J.}\ \bibnamefont {Humphreys}}, \ and\ \bibinfo
  {author} {\bibfnamefont {A.~P.}\ \bibnamefont {Sutton}},\ }\href {\doibase
  10.1103/PhysRevB.57.1505} {\bibfield  {journal} {\bibinfo  {journal}
  {Physical Review B}\ }\textbf {\bibinfo {volume} {57}},\ \bibinfo {pages}
  {1505} (\bibinfo {year} {1998})}\BibitemShut {NoStop}%
\bibitem [{\citenamefont {Blume}\ and\ \citenamefont
  {Hsieh}(1969)}]{Blume_Hsieh}%
  \BibitemOpen
  \bibfield  {author} {\bibinfo {author} {\bibfnamefont {M.}~\bibnamefont
  {Blume}}\ and\ \bibinfo {author} {\bibfnamefont {Y.~Y.}\ \bibnamefont
  {Hsieh}},\ }\href {\doibase 10.1063/1.1657616} {\bibfield  {journal}
  {\bibinfo  {journal} {Journal of Applied Physics}\ }\textbf {\bibinfo
  {volume} {40}},\ \bibinfo {pages} {1249} (\bibinfo {year} {1969})},\ \Eprint
  {http://arxiv.org/abs/https://doi.org/10.1063/1.1657616}
  {https://doi.org/10.1063/1.1657616} \BibitemShut {NoStop}%
\bibitem [{\citenamefont {Bhatt}\ and\ \citenamefont
  {Yang}(1998)}]{Bhatt_Yang}%
  \BibitemOpen
  \bibfield  {author} {\bibinfo {author} {\bibfnamefont {R.~N.}\ \bibnamefont
  {Bhatt}}\ and\ \bibinfo {author} {\bibfnamefont {K.}~\bibnamefont {Yang}},\
  }\href {\doibase 10.1063/1.367612} {\bibfield  {journal} {\bibinfo  {journal}
  {Journal of Applied Physics}\ }\textbf {\bibinfo {volume} {83}},\ \bibinfo
  {pages} {7231} (\bibinfo {year} {1998})},\ \Eprint
  {http://arxiv.org/abs/https://doi.org/10.1063/1.367612}
  {https://doi.org/10.1063/1.367612} \BibitemShut {NoStop}%
\bibitem [{\citenamefont {{Mila}}\ and\ \citenamefont
  {{Zhang}}(2000)}]{Mila_Zhang}%
  \BibitemOpen
  \bibfield  {author} {\bibinfo {author} {\bibfnamefont {F.}~\bibnamefont
  {{Mila}}}\ and\ \bibinfo {author} {\bibfnamefont {F.-C.}\ \bibnamefont
  {{Zhang}}},\ }\href {\doibase 10.1007/s100510070242} {\bibfield  {journal}
  {\bibinfo  {journal} {European Physical Journal B}\ }\textbf {\bibinfo
  {volume} {16}},\ \bibinfo {pages} {7} (\bibinfo {year} {2000})},\ \Eprint
  {http://arxiv.org/abs/cond-mat/0006068} {arXiv:cond-mat/0006068
  [cond-mat.str-el]} \BibitemShut {NoStop}%
\bibitem [{\citenamefont {Fazekas}(1999)}]{fazekas1999}%
  \BibitemOpen
  \bibfield  {author} {\bibinfo {author} {\bibfnamefont {P.}~\bibnamefont
  {Fazekas}},\ }\href@noop {} {\emph {\bibinfo {title} {Lecture notes on
  electron correlation and magnetism}}},\ Vol.~\bibinfo {volume} {5}\ (\bibinfo
   {publisher} {World scientific},\ \bibinfo {year} {1999})\BibitemShut
  {NoStop}%
\bibitem [{\citenamefont {{Desai}}\ and\ \citenamefont
  {{Kaul}}(2019)}]{Desai_Kaul}%
  \BibitemOpen
  \bibfield  {author} {\bibinfo {author} {\bibfnamefont {N.}~\bibnamefont
  {{Desai}}}\ and\ \bibinfo {author} {\bibfnamefont {R.}~\bibnamefont
  {{Kaul}}},\ }\href@noop {} {\bibfield  {journal} {\bibinfo  {journal} {arXiv
  e-prints}\ ,\ \bibinfo {eid} {arXiv:1904.09629}} (\bibinfo {year} {2019})},\
  \Eprint {http://arxiv.org/abs/1904.09629} {arXiv:1904.09629
  [cond-mat.str-el]} \BibitemShut {NoStop}%
\bibitem [{\citenamefont {Oitmaa}\ and\ \citenamefont
  {Singh}(2016)}]{Oitmaa_Singh}%
  \BibitemOpen
  \bibfield  {author} {\bibinfo {author} {\bibfnamefont {J.}~\bibnamefont
  {Oitmaa}}\ and\ \bibinfo {author} {\bibfnamefont {R.~R.~P.}\ \bibnamefont
  {Singh}},\ }\href {\doibase 10.1103/PhysRevB.93.014424} {\bibfield  {journal}
  {\bibinfo  {journal} {Phys. Rev. B}\ }\textbf {\bibinfo {volume} {93}},\
  \bibinfo {pages} {014424} (\bibinfo {year} {2016})}\BibitemShut {NoStop}%
\bibitem [{\citenamefont {Cohen}(1992)}]{Cohen1992}%
  \BibitemOpen
  \bibfield  {author} {\bibinfo {author} {\bibfnamefont {R.~E.}\ \bibnamefont
  {Cohen}},\ }\href {\doibase 10.1038/358136a0} {\bibfield  {journal} {\bibinfo
   {journal} {Nature}\ }\textbf {\bibinfo {volume} {358}},\ \bibinfo {pages}
  {136} (\bibinfo {year} {1992})}\BibitemShut {NoStop}%
\bibitem [{\citenamefont {Birol}\ \emph {et~al.}(2012)\citenamefont {Birol},
  \citenamefont {Benedek}, \citenamefont {Das}, \citenamefont {Wysocki},
  \citenamefont {Mulder}, \citenamefont {Abbett}, \citenamefont {Smith},
  \citenamefont {Ghosh},\ and\ \citenamefont {Fennie}}]{Birol2012}%
  \BibitemOpen
  \bibfield  {author} {\bibinfo {author} {\bibfnamefont {T.}~\bibnamefont
  {Birol}}, \bibinfo {author} {\bibfnamefont {N.~A.}\ \bibnamefont {Benedek}},
  \bibinfo {author} {\bibfnamefont {H.}~\bibnamefont {Das}}, \bibinfo {author}
  {\bibfnamefont {A.~L.}\ \bibnamefont {Wysocki}}, \bibinfo {author}
  {\bibfnamefont {A.~T.}\ \bibnamefont {Mulder}}, \bibinfo {author}
  {\bibfnamefont {B.~M.}\ \bibnamefont {Abbett}}, \bibinfo {author}
  {\bibfnamefont {E.~H.}\ \bibnamefont {Smith}}, \bibinfo {author}
  {\bibfnamefont {S.}~\bibnamefont {Ghosh}}, \ and\ \bibinfo {author}
  {\bibfnamefont {C.~J.}\ \bibnamefont {Fennie}},\ }\href {\doibase
  10.1016/j.cossms.2012.08.002} {\bibfield  {journal} {\bibinfo  {journal}
  {Current Opinion in Solid State and Materials Science}\ }\textbf {\bibinfo
  {volume} {16}},\ \bibinfo {pages} {227} (\bibinfo {year} {2012})}\BibitemShut
  {NoStop}%
\bibitem [{\citenamefont {Wysocki}\ and\ \citenamefont
  {Birol}(2016)}]{Wysocki2016}%
  \BibitemOpen
  \bibfield  {author} {\bibinfo {author} {\bibfnamefont {A.~L.}\ \bibnamefont
  {Wysocki}}\ and\ \bibinfo {author} {\bibfnamefont {T.}~\bibnamefont
  {Birol}},\ }\href {\doibase 10.1103/PhysRevB.93.134425} {\bibfield  {journal}
  {\bibinfo  {journal} {Physical Review B}\ }\textbf {\bibinfo {volume} {93}},\
  \bibinfo {pages} {134425} (\bibinfo {year} {2016})},\ \Eprint
  {http://arxiv.org/abs/arXiv:1508.00834v1} {arXiv:1508.00834v1} \BibitemShut
  {NoStop}%
\bibitem [{\citenamefont {Zylbersztejn}\ and\ \citenamefont
  {Mott}(1975)}]{Zylbersztejn1975}%
  \BibitemOpen
  \bibfield  {author} {\bibinfo {author} {\bibfnamefont {A.}~\bibnamefont
  {Zylbersztejn}}\ and\ \bibinfo {author} {\bibfnamefont {N.~F.}\ \bibnamefont
  {Mott}},\ }\href {\doibase 10.1103/PhysRevB.11.4383} {\bibfield  {journal}
  {\bibinfo  {journal} {Physical Review B}\ }\textbf {\bibinfo {volume} {11}},\
  \bibinfo {pages} {4383} (\bibinfo {year} {1975})}\BibitemShut {NoStop}%
\bibitem [{\citenamefont {Wentzcovitch}\ \emph {et~al.}(1994)\citenamefont
  {Wentzcovitch}, \citenamefont {Schulz},\ and\ \citenamefont
  {Allen}}]{Wentzcovitch1994}%
  \BibitemOpen
  \bibfield  {author} {\bibinfo {author} {\bibfnamefont {R.~M.}\ \bibnamefont
  {Wentzcovitch}}, \bibinfo {author} {\bibfnamefont {W.~W.}\ \bibnamefont
  {Schulz}}, \ and\ \bibinfo {author} {\bibfnamefont {P.~B.}\ \bibnamefont
  {Allen}},\ }\href {\doibase 10.1103/PhysRevLett.72.3389} {\bibfield
  {journal} {\bibinfo  {journal} {Physical Review Letters}\ }\textbf {\bibinfo
  {volume} {72}},\ \bibinfo {pages} {3389} (\bibinfo {year}
  {1994})}\BibitemShut {NoStop}%
\bibitem [{\citenamefont {Haverkort}\ \emph {et~al.}(2005)\citenamefont
  {Haverkort}, \citenamefont {Hu}, \citenamefont {Tanaka}, \citenamefont
  {Reichelt}, \citenamefont {Streltsov}, \citenamefont {Korotin}, \citenamefont
  {Anisimov}, \citenamefont {Hsieh}, \citenamefont {Lin}, \citenamefont {Chen},
  \citenamefont {Khomskii},\ and\ \citenamefont {Tjeng}}]{Haverkort2005}%
  \BibitemOpen
  \bibfield  {author} {\bibinfo {author} {\bibfnamefont {M.~W.}\ \bibnamefont
  {Haverkort}}, \bibinfo {author} {\bibfnamefont {Z.}~\bibnamefont {Hu}},
  \bibinfo {author} {\bibfnamefont {A.}~\bibnamefont {Tanaka}}, \bibinfo
  {author} {\bibfnamefont {W.}~\bibnamefont {Reichelt}}, \bibinfo {author}
  {\bibfnamefont {S.~V.}\ \bibnamefont {Streltsov}}, \bibinfo {author}
  {\bibfnamefont {M.~A.}\ \bibnamefont {Korotin}}, \bibinfo {author}
  {\bibfnamefont {V.~I.}\ \bibnamefont {Anisimov}}, \bibinfo {author}
  {\bibfnamefont {H.~H.}\ \bibnamefont {Hsieh}}, \bibinfo {author}
  {\bibfnamefont {H.-J.}\ \bibnamefont {Lin}}, \bibinfo {author} {\bibfnamefont
  {C.~T.}\ \bibnamefont {Chen}}, \bibinfo {author} {\bibfnamefont {D.~I.}\
  \bibnamefont {Khomskii}}, \ and\ \bibinfo {author} {\bibfnamefont {L.~H.}\
  \bibnamefont {Tjeng}},\ }\href {\doibase 10.1103/PhysRevLett.95.196404}
  {\bibfield  {journal} {\bibinfo  {journal} {Physical Review Letters}\
  }\textbf {\bibinfo {volume} {95}},\ \bibinfo {pages} {196404} (\bibinfo
  {year} {2005})}\BibitemShut {NoStop}%
\bibitem [{\citenamefont {Biermann}\ \emph {et~al.}(2005)\citenamefont
  {Biermann}, \citenamefont {Poteryaev}, \citenamefont {Lichtenstein},\ and\
  \citenamefont {Georges}}]{Biermann2005}%
  \BibitemOpen
  \bibfield  {author} {\bibinfo {author} {\bibfnamefont {S.}~\bibnamefont
  {Biermann}}, \bibinfo {author} {\bibfnamefont {A.}~\bibnamefont {Poteryaev}},
  \bibinfo {author} {\bibfnamefont {A.~I.}\ \bibnamefont {Lichtenstein}}, \
  and\ \bibinfo {author} {\bibfnamefont {A.}~\bibnamefont {Georges}},\ }\href
  {\doibase 10.1103/PhysRevLett.94.026404} {\bibfield  {journal} {\bibinfo
  {journal} {Physical Review Letters}\ }\textbf {\bibinfo {volume} {94}},\
  \bibinfo {pages} {026404} (\bibinfo {year} {2005})}\BibitemShut {NoStop}%
\bibitem [{\citenamefont {Weber}\ \emph {et~al.}(2012)\citenamefont {Weber},
  \citenamefont {O'Regan}, \citenamefont {Hine}, \citenamefont {Payne},
  \citenamefont {Kotliar},\ and\ \citenamefont {Littlewood}}]{Weber2012}%
  \BibitemOpen
  \bibfield  {author} {\bibinfo {author} {\bibfnamefont {C.}~\bibnamefont
  {Weber}}, \bibinfo {author} {\bibfnamefont {D.~D.}\ \bibnamefont {O'Regan}},
  \bibinfo {author} {\bibfnamefont {N.~D.~M.}\ \bibnamefont {Hine}}, \bibinfo
  {author} {\bibfnamefont {M.~C.}\ \bibnamefont {Payne}}, \bibinfo {author}
  {\bibfnamefont {G.}~\bibnamefont {Kotliar}}, \ and\ \bibinfo {author}
  {\bibfnamefont {P.~B.}\ \bibnamefont {Littlewood}},\ }\href {\doibase
  10.1103/PhysRevLett.108.256402} {\bibfield  {journal} {\bibinfo  {journal}
  {Physical Review Letters}\ }\textbf {\bibinfo {volume} {108}},\ \bibinfo
  {pages} {256402} (\bibinfo {year} {2012})}\BibitemShut {NoStop}%
\bibitem [{\citenamefont {Gray}\ \emph {et~al.}(2016)\citenamefont {Gray},
  \citenamefont {Jeong}, \citenamefont {Aetukuri}, \citenamefont {Granitzka},
  \citenamefont {Chen}, \citenamefont {Kukreja}, \citenamefont {Higley},
  \citenamefont {Chase}, \citenamefont {Reid}, \citenamefont {Ohldag},
  \citenamefont {Marcus}, \citenamefont {Scholl}, \citenamefont {Young},
  \citenamefont {Doran}, \citenamefont {Jenkins}, \citenamefont {Shafer},
  \citenamefont {Arenholz}, \citenamefont {Samant}, \citenamefont {Parkin},\
  and\ \citenamefont {D{\"{u}}rr}}]{Gray2016}%
  \BibitemOpen
  \bibfield  {author} {\bibinfo {author} {\bibfnamefont {A.~X.}\ \bibnamefont
  {Gray}}, \bibinfo {author} {\bibfnamefont {J.}~\bibnamefont {Jeong}},
  \bibinfo {author} {\bibfnamefont {N.~P.}\ \bibnamefont {Aetukuri}}, \bibinfo
  {author} {\bibfnamefont {P.}~\bibnamefont {Granitzka}}, \bibinfo {author}
  {\bibfnamefont {Z.}~\bibnamefont {Chen}}, \bibinfo {author} {\bibfnamefont
  {R.}~\bibnamefont {Kukreja}}, \bibinfo {author} {\bibfnamefont
  {D.}~\bibnamefont {Higley}}, \bibinfo {author} {\bibfnamefont
  {T.}~\bibnamefont {Chase}}, \bibinfo {author} {\bibfnamefont {A.~H.}\
  \bibnamefont {Reid}}, \bibinfo {author} {\bibfnamefont {H.}~\bibnamefont
  {Ohldag}}, \bibinfo {author} {\bibfnamefont {M.~A.}\ \bibnamefont {Marcus}},
  \bibinfo {author} {\bibfnamefont {A.}~\bibnamefont {Scholl}}, \bibinfo
  {author} {\bibfnamefont {A.~T.}\ \bibnamefont {Young}}, \bibinfo {author}
  {\bibfnamefont {A.}~\bibnamefont {Doran}}, \bibinfo {author} {\bibfnamefont
  {C.~A.}\ \bibnamefont {Jenkins}}, \bibinfo {author} {\bibfnamefont
  {P.}~\bibnamefont {Shafer}}, \bibinfo {author} {\bibfnamefont
  {E.}~\bibnamefont {Arenholz}}, \bibinfo {author} {\bibfnamefont {M.~G.}\
  \bibnamefont {Samant}}, \bibinfo {author} {\bibfnamefont {S.~S.~P.}\
  \bibnamefont {Parkin}}, \ and\ \bibinfo {author} {\bibfnamefont {H.~A.}\
  \bibnamefont {D{\"{u}}rr}},\ }\href {\doibase 10.1103/PhysRevLett.116.116403}
  {\bibfield  {journal} {\bibinfo  {journal} {Physical Review Letters}\
  }\textbf {\bibinfo {volume} {116}},\ \bibinfo {pages} {116403} (\bibinfo
  {year} {2016})}\BibitemShut {NoStop}%
\bibitem [{\citenamefont {Quackenbush}\ \emph {et~al.}(2013)\citenamefont
  {Quackenbush}, \citenamefont {Tashman}, \citenamefont {Mundy}, \citenamefont
  {Sallis}, \citenamefont {Paik}, \citenamefont {Misra}, \citenamefont {Moyer},
  \citenamefont {Guo}, \citenamefont {Fischer}, \citenamefont {Woicik},
  \citenamefont {Muller}, \citenamefont {Schlom},\ and\ \citenamefont
  {Piper}}]{Quackenbush2013}%
  \BibitemOpen
  \bibfield  {author} {\bibinfo {author} {\bibfnamefont {N.~F.}\ \bibnamefont
  {Quackenbush}}, \bibinfo {author} {\bibfnamefont {J.~W.}\ \bibnamefont
  {Tashman}}, \bibinfo {author} {\bibfnamefont {J.~A.}\ \bibnamefont {Mundy}},
  \bibinfo {author} {\bibfnamefont {S.}~\bibnamefont {Sallis}}, \bibinfo
  {author} {\bibfnamefont {H.}~\bibnamefont {Paik}}, \bibinfo {author}
  {\bibfnamefont {R.}~\bibnamefont {Misra}}, \bibinfo {author} {\bibfnamefont
  {J.~A.}\ \bibnamefont {Moyer}}, \bibinfo {author} {\bibfnamefont {J.-H.}\
  \bibnamefont {Guo}}, \bibinfo {author} {\bibfnamefont {D.~A.}\ \bibnamefont
  {Fischer}}, \bibinfo {author} {\bibfnamefont {J.~C.}\ \bibnamefont {Woicik}},
  \bibinfo {author} {\bibfnamefont {D.~A.}\ \bibnamefont {Muller}}, \bibinfo
  {author} {\bibfnamefont {D.~G.}\ \bibnamefont {Schlom}}, \ and\ \bibinfo
  {author} {\bibfnamefont {L.~F.~J.}\ \bibnamefont {Piper}},\ }\href {\doibase
  10.1021/nl402716d} {\bibfield  {journal} {\bibinfo  {journal} {Nano Letters}\
  }\textbf {\bibinfo {volume} {13}},\ \bibinfo {pages} {4857} (\bibinfo {year}
  {2013})}\BibitemShut {NoStop}%
\bibitem [{\citenamefont {Huffman}\ \emph {et~al.}(2017)\citenamefont
  {Huffman}, \citenamefont {Hendriks}, \citenamefont {Walter}, \citenamefont
  {Yoon}, \citenamefont {Ju}, \citenamefont {Smith}, \citenamefont {Carr},
  \citenamefont {Krakauer},\ and\ \citenamefont {Qazilbash}}]{Huffman2017}%
  \BibitemOpen
  \bibfield  {author} {\bibinfo {author} {\bibfnamefont {T.~J.}\ \bibnamefont
  {Huffman}}, \bibinfo {author} {\bibfnamefont {C.}~\bibnamefont {Hendriks}},
  \bibinfo {author} {\bibfnamefont {E.~J.}\ \bibnamefont {Walter}}, \bibinfo
  {author} {\bibfnamefont {J.}~\bibnamefont {Yoon}}, \bibinfo {author}
  {\bibfnamefont {H.}~\bibnamefont {Ju}}, \bibinfo {author} {\bibfnamefont
  {R.}~\bibnamefont {Smith}}, \bibinfo {author} {\bibfnamefont {G.~L.}\
  \bibnamefont {Carr}}, \bibinfo {author} {\bibfnamefont {H.}~\bibnamefont
  {Krakauer}}, \ and\ \bibinfo {author} {\bibfnamefont {M.~M.}\ \bibnamefont
  {Qazilbash}},\ }\href {\doibase 10.1103/PhysRevB.95.075125} {\bibfield
  {journal} {\bibinfo  {journal} {Physical Review B}\ }\textbf {\bibinfo
  {volume} {95}},\ \bibinfo {pages} {075125} (\bibinfo {year}
  {2017})}\BibitemShut {NoStop}%
\bibitem [{\citenamefont {N\'ajera}\ \emph {et~al.}(2017)\citenamefont
  {N\'ajera}, \citenamefont {Civelli}, \citenamefont
  {Dobrosavljevi\ifmmode~\acute{c}\else \'{c}\fi{}},\ and\ \citenamefont
  {Rozenberg}}]{Vlad_VO2}%
  \BibitemOpen
  \bibfield  {author} {\bibinfo {author} {\bibfnamefont {O.}~\bibnamefont
  {N\'ajera}}, \bibinfo {author} {\bibfnamefont {M.}~\bibnamefont {Civelli}},
  \bibinfo {author} {\bibfnamefont {V.}~\bibnamefont
  {Dobrosavljevi\ifmmode~\acute{c}\else \'{c}\fi{}}}, \ and\ \bibinfo {author}
  {\bibfnamefont {M.~J.}\ \bibnamefont {Rozenberg}},\ }\href {\doibase
  10.1103/PhysRevB.95.035113} {\bibfield  {journal} {\bibinfo  {journal} {Phys.
  Rev. B}\ }\textbf {\bibinfo {volume} {95}},\ \bibinfo {pages} {035113}
  (\bibinfo {year} {2017})}\BibitemShut {NoStop}%
\bibitem [{ITe()}]{ITensor}%
  \BibitemOpen
  \href@noop {} {}\bibinfo {note} {M. Stoudenmire and S.R. White,
  www.itensor.org}\BibitemShut {NoStop}%
\bibitem [{\citenamefont {Quenouille}(1949)}]{quenouille1949}%
  \BibitemOpen
  \bibfield  {author} {\bibinfo {author} {\bibfnamefont {M.~H.}\ \bibnamefont
  {Quenouille}},\ }\href@noop {} {\bibfield  {journal} {\bibinfo  {journal}
  {Ann. Math. Statist.}\ }\textbf {\bibinfo {volume} {20}},\ \bibinfo {pages}
  {355} (\bibinfo {year} {1949})}\BibitemShut {NoStop}%
\bibitem [{\citenamefont {QUENOUILLE}(1956)}]{quenouille1956}%
  \BibitemOpen
  \bibfield  {author} {\bibinfo {author} {\bibfnamefont {M.~H.}\ \bibnamefont
  {QUENOUILLE}},\ }\href@noop {} {\bibfield  {journal} {\bibinfo  {journal}
  {Biometrika}\ }\textbf {\bibinfo {volume} {43}},\ \bibinfo {pages} {353}
  (\bibinfo {year} {1956})}\BibitemShut {NoStop}%
\bibitem [{\citenamefont {Tukey}(1958)}]{Tukey1958}%
  \BibitemOpen
  \bibfield  {author} {\bibinfo {author} {\bibfnamefont {J.~W.}\ \bibnamefont
  {Tukey}},\ }\href@noop {} {\bibfield  {journal} {\bibinfo  {journal} {Ann.
  Math. Statist.}\ }\textbf {\bibinfo {volume} {29}},\ \bibinfo {pages} {614}
  (\bibinfo {year} {1958})}\BibitemShut {NoStop}%
\end{thebibliography}%

\clearpage
\newpage
\beginsupplement
%%%%%%%%%% Merge with supplemental materials %%%%%%%%%%
\widetext
\begin{center}
\Large{\textbf{Supplementary Information: Spin--lattice coupling and the emergence of the trimerized phase in the $S=1$ Kagome antiferromagnet Na$_2$Ti$_3$Cl$_8$}}
\end{center}

%%%%%%%%%% Merge with supplemental materials %%%%%%%%%%
%%%%%%%%%% Prefix a "S" to all equations, figures, tables and reset the counter %%%%%%%%%%
\setcounter{equation}{0}
\setcounter{table}{0}
\setcounter{page}{1}
\makeatletter
\renewcommand{\theequation}{S\arabic{equation}}
\renewcommand{\thefigure}{S\arabic{figure}}
\renewcommand{\bibnumfmt}[1]{[S#1]}
\renewcommand{\citenumfont}[1]{S#1}

\section{Magnetic Hamiltonian for three sites}
\begin{figure}[!h]
\centering%
		\includegraphics[width=0.5\textwidth]{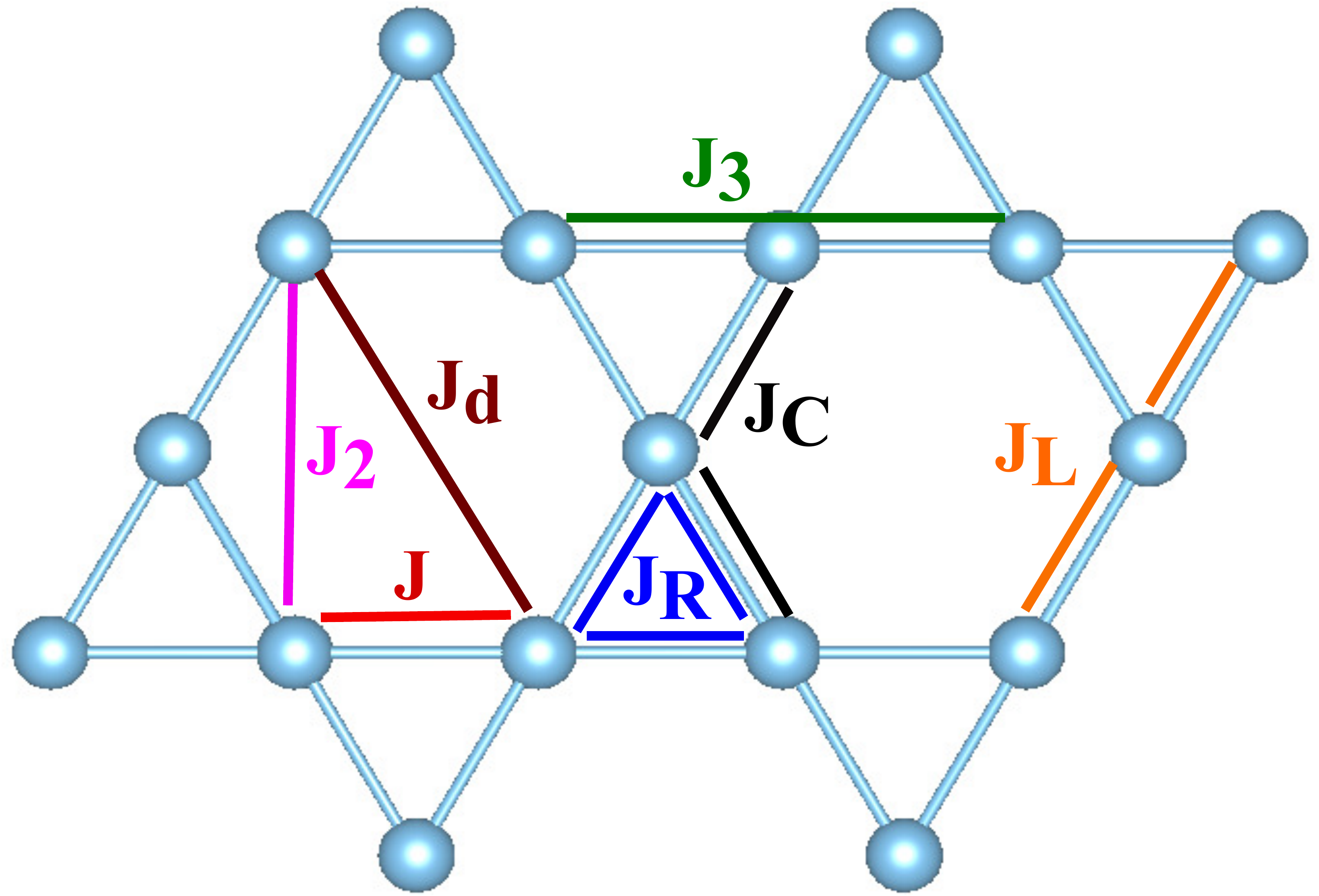}%
		\caption{$J$ (nearest neighbor), $J_2$ (second nearest neighbor), $J_3$ and $J_d$ (third nearest neighbor), $J_R$ (intra-triangle ring exchange) and $J_L$ (inter-triangle interaction) denote different paths of magnetic exchange interactions.}%
	\label{FigS1}%
\end{figure}

Along with the conventional three-site intra-triangle ring exchange ($J_R$, see Fig.~\ref{FigS1}), there are also two distinct symmetrically allowed exchange terms ($J_C$ and $J_L$, see Fig.~\ref{FigS1}) that involves interactions of spins between the two corner sharing triangles. The model Hamiltonian that includes second (bilinear, nearest neighbor) and fourth order (biquadratic and three-site exchange) coupling of nearest neighbor spins is given as,
\begin{equation}
\begin{split}
H_{bl-bq-three-spin} & =J\sum_{\langle ij \rangle}\hat{s}_{i}.\hat{s}_{j}+J_{bq}\sum_{\langle ij \rangle}(\hat{s}_{i}.\hat{s}_{j})^{2}+\frac{J_{R}}{2}\sum_{\bigtriangleup = i,j,k}((\hat{s}_{i}.\hat{s}_{j})(\hat{s}_{i}.\hat{s}_{k})+(\hat{s}_{i}.\hat{s}_{k})(\hat{s}_{i}.\hat{s}_{j}))\\
& +\frac{J_{L}}{2}\sum_{\triangleright\triangleleft}((\hat{s}_{i}.\hat{s}_{j})(\hat{s}_{i}.\hat{s}_{k})+(\hat{s}_{i}.\hat{s}_{k})(\hat{s}_{i}.\hat{s}_{j}))+ \frac{J_{C}}{2}\sum_{\triangleright\triangleleft}((\hat{s}_{i}.\hat{s}_{j})(\hat{s}_{i}.\hat{s}_{k})+(\hat{s}_{i}.\hat{s}_{k})(\hat{s}_{i}.\hat{s}_{j})).
\end{split}
\end{equation}
$J_L$ (inter-triangle, see Fig.~\ref{FigS1}) and $J_C$ (inter-triangle, see Fig.~\ref{FigS1}) denote fourth order coupling constants arising from the interaction of spins at three different lattice sites similar to $J_R$. We note that all of these terms are symmetry allowed in the absence of spin orbit coupling, and require hopping only between nearest neighbor Ti atoms. 

\begin{figure}[!h]
	\centering%
		\includegraphics[width=0.95\textwidth]{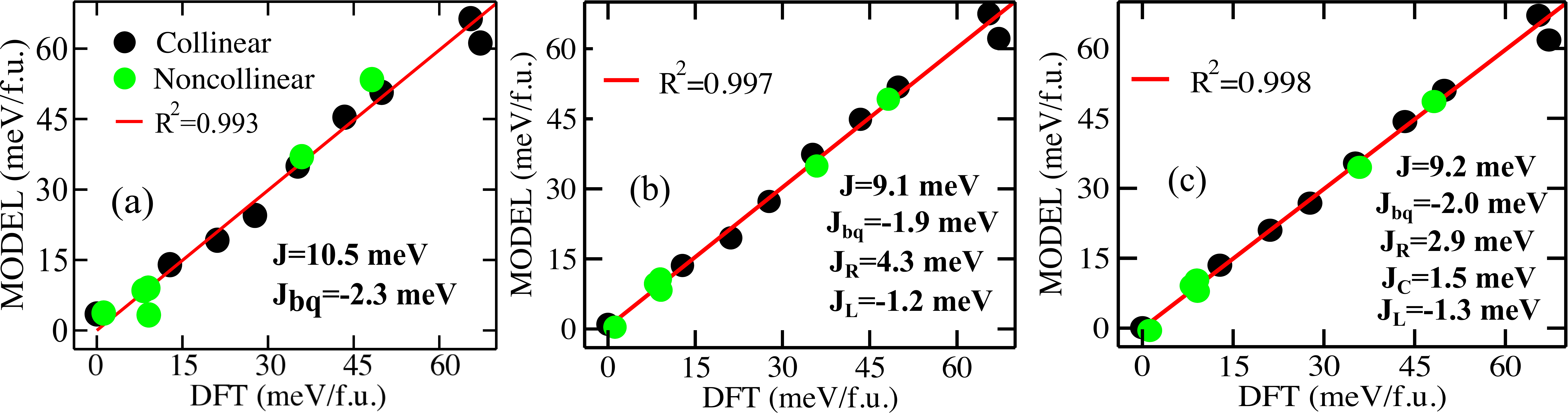}%
		\caption{Fitting of energies ($U$=3 eV) of several magnetic states with bilinear-biquadratic-three spin exchange model with (a) four ($J$, $J_{bq}$, $J_R$ and $J_L$) and (b) five ($J$, $J_{bq}$, $J_R$, $J_L$ and $J_C$) coupling parameters.}%
	\label{FigS2}%
\end{figure}%
\begin{figure}
	\centering%
		\includegraphics[width=0.99\textwidth]{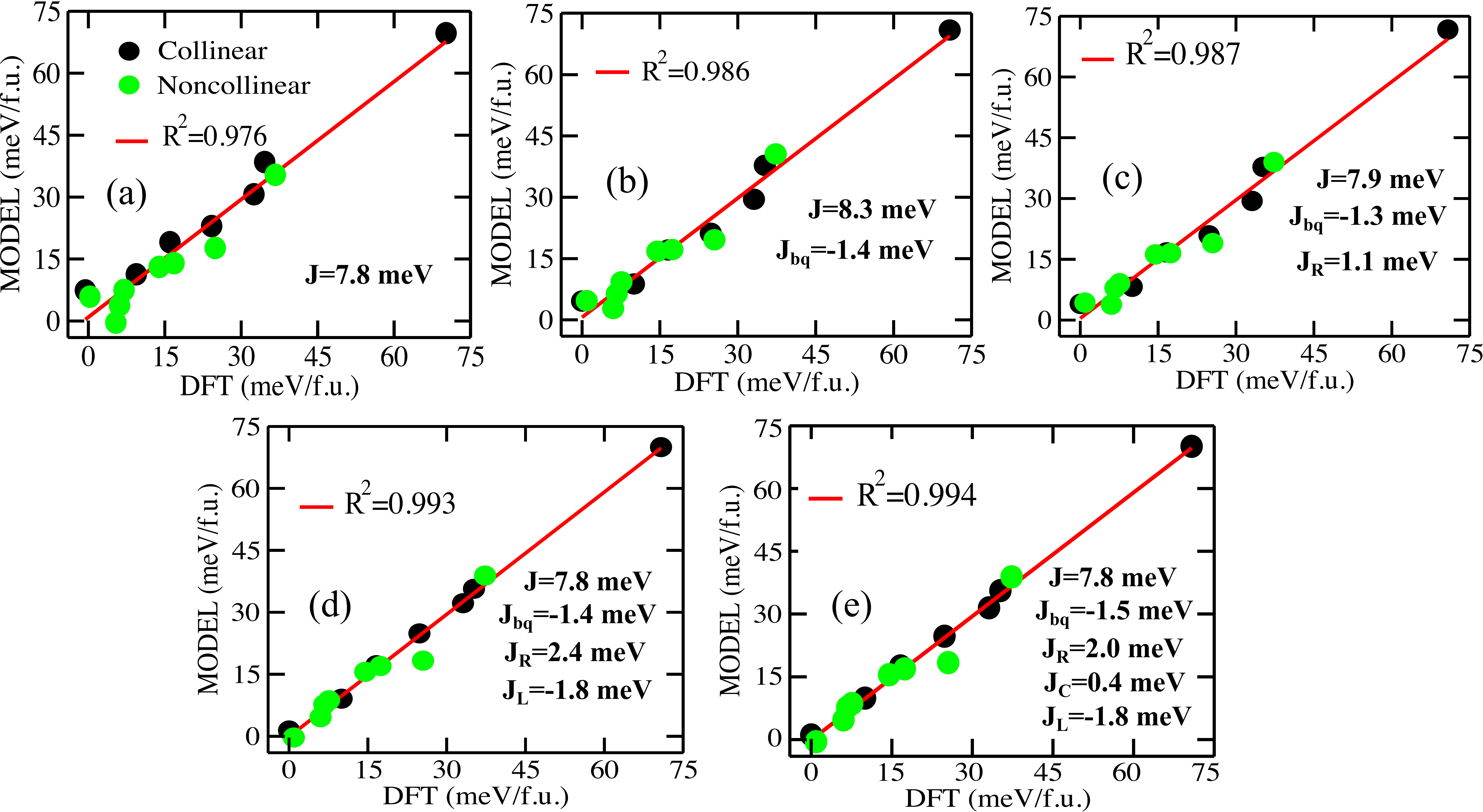}%
		\caption{Fitting of energies ($U$=4 eV) of several magnetic states with bilinear-biquadratic-three spin exchange model with (a) one ($J$), (b) two ($J$ and $J_{bq}$), (c) three ($J$, $J_{bq}$ and $J_R$), (d) four ($J$, $J_{bq}$, $J_R$ and $J_L$) and (b) five ($J$, $J_{bq}$, $J_R$, $J_L$ and $J_C$) parameters.}%
	\label{FigS3}%
\end{figure}%
To determine the values of five different coupling parameters ($J$, $J_{bq}$, $J_C$, $J_R$, $J_L$), we fit energies of several collinear and non-collinear magnetic configurations. When the energies are fitted to the Hamiltonian involving four ($J$, $J_{bq}$, $J_R$, $J_L$) or five ($J$, $J_{bq}$, $J_R$, $J_L$, $J_C$) coupling parameters, $J$ and $J_{bq}$ change by 0.8\% and 15\% from their values with three parameter model or the bilinear-biquadratic-ring exchange model ($J$, $J_{bq}$, $J_R$, see Fig.~2 
in the main manuscript, Fig.~\ref{FigS2} and Fig. \ref{FigS3}). Including $J_L$ and $J_C$ parameters improve the statistics of fitting as evident in the values of $R^2$ (see Fig. \ref{FigS2}). This highlights that model which includes three site exchange term is more superior than the bilinear-biquadratic model (see Fig. \ref{FigS2} and Fig. \ref{FigS3}). 

$J_L$ prefers the spins at the two neighboring sites (inter-triangle) to be  either parallel or anti-parallel to its spin (see Fig. \ref{FigS1}-\ref{FigS3}). On the other hand, $J_C$ forces spins at the two neighboring sites to be parallel to its spin. $J_C$ and $J_L$ are slightly smaller compared to $J_R$ in the high temperature phase. 
$J_L$ and $J_C$ are expected to be negligible in the low temperature phase as the hopping of electrons between neighboring Ti sites inside the large triangle decreases drastically (see Fig. 3 in main manuscript) as a function of structural distortion from the high temperature phase. 

\section{Bilinear-biquadratic model}

In order to compare the fits we have obtained using the models presented so far, we also consider another Hamiltonian which consists of nearest neighbor biquadratic coupling in addition to bilinear couplings for further neighbors. 
\begin{equation}
H_{bl-bq}=J\sum_{ij, nn}\hat{s}_{i}.\hat{s}_{j}+J_{2}\sum_{ij, 2nn}\hat{s}_{i}.\hat{s}_{j}+J_{3}\sum_{ij,3nn}\hat{s}_{i}.\hat{s}_{j}+J_{d}\sum_{ij,3nn}\hat{s}_{i}.\hat{s}_{j}+J_{bq}\sum_{ij, nn}(\hat{s}_{i}.\hat{s}_{j})^{2}.
\end{equation}
\begin{figure}
	\centering%
		\includegraphics[width=0.95\textwidth]{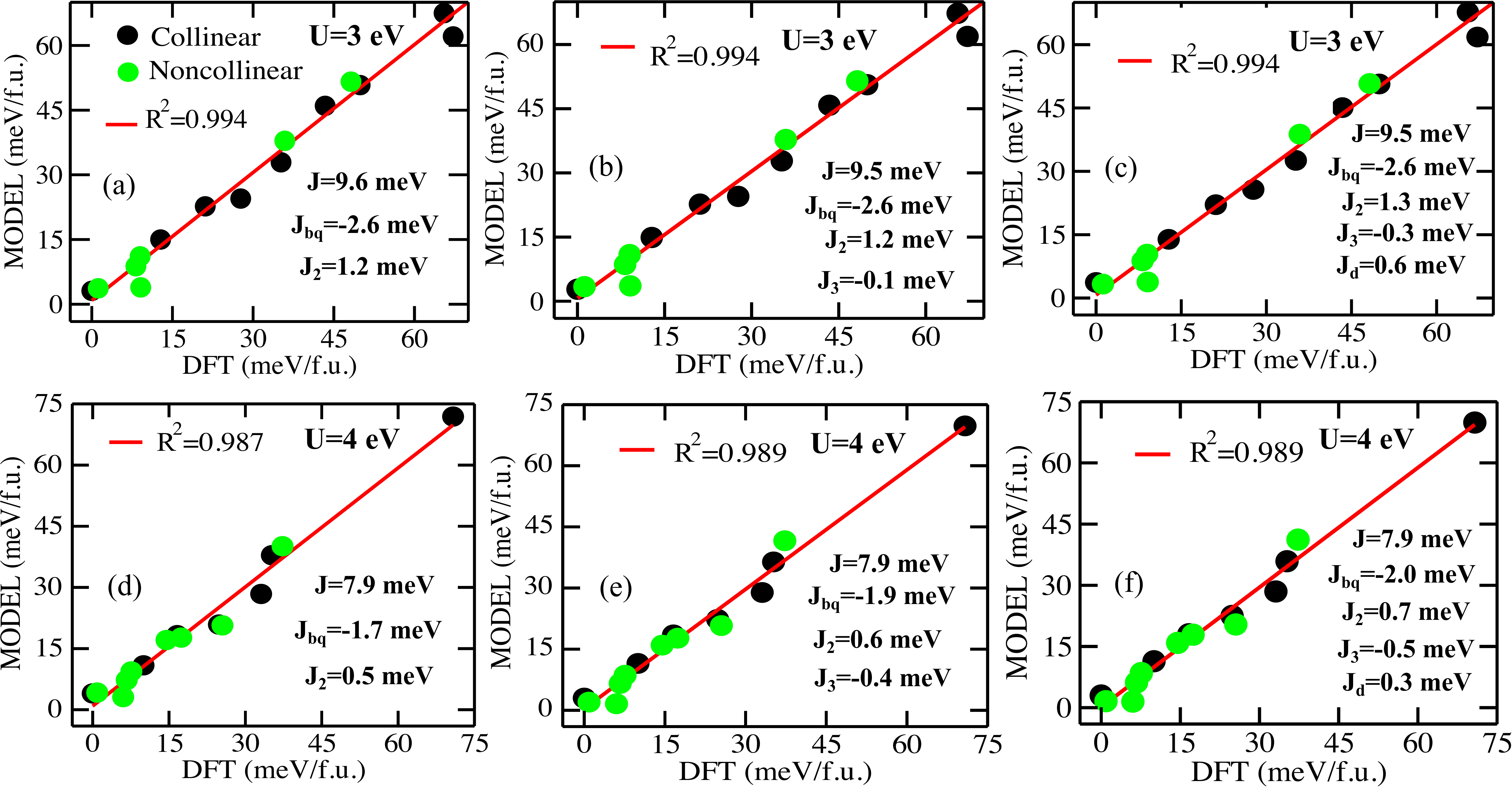}%
		\caption{Fitting of energies of several magnetic states with bilinear-biquadratic model with three ($J$, $J_{bq}$ and $J_2$), (d) four ($J$, $J_{bq}$, $J_2$ and $J_3$) and (b) five ($J$, $J_{bq}$, $J_2$, $J_3$ and $J_d$) parameters.}%
	\label{FigS4}%
\end{figure}%
(Some of these further neighbor couplings were shown, from first principles, to be important in the $S$=1/2 kagome magnet Herbertsmithite\cite{Jeschke_Valenti}.)
This model includes bilinear exchange of spins up to third nearest neighbors (second neighbor: $J_2$ and third neighbor: $J_3$ and $J_d$, see Fig. \ref{FigS1}) along with the nearest neighbor bilinear ($J$) and biquadratic ($J_{bq}$) exchanges. The biquadratic exchange term is linearly dependent on the bilinear term for spin-1/2 Hamiltonians 
and hence not applicable to Herbertsmithite. We consider this term in the model as it is important for systems with $S$=1 localized spins. This model also predicts 
the nearest neighbor bilinear exchange to be antiferromagnetic and biquadratic exchange to favor the collinear spin configurations similar to bilinear-biquadratic-three spin exchange model (see Fig. \ref{FigS4} and Fig. \ref{FigS2}-\ref{FigS3}). Estimate of nearest neighbor bilinear coupling constant ($J$) using this model slightly differs from its value calculated from the bilinear-biquadratic-three spin exchange model (see Fig. \ref{FigS4} and Fig. \ref{FigS2}-\ref{FigS3}). The biquadratic coupling constant increases by 30-47 \% compared to its value estimated using bilinear-biquadratic-three spin exchange model. 

$J_2$ (second nearest neighbor) and $J_d$ (third nearest neighbor) prefer antiferromagnetic alignment of spins. In contrast, $J_3$ (third nearest neighbor) is ferromagnetic in nature (see Fig.~\ref{FigS4}). Our estimated values of $J_2$, $J_3$ and $J_d$ are negligible (see Fig. \ref{FigS4}) compared to the nearest neighbor bilinear coupling constant ($J$) as the Ti-Ti distance between second and third nearest neighbors are large (Ti-Ti (second nn)=6.40 $\angstrom$ and Ti-Ti (third nn)=7.39 $\angstrom$). In addition, we find that fitting of energies has bad statistics (small $R^2$) compared to the bilinear-biquadratic-three spin exchange model (Fig. \ref{FigS4}), and it highlights the superiority of the later one. 

\section{Jackknife resampling}

Jackknife resampling is a resampling technique similar to bootstrap and permutation test methods\cite{quenouille1949,quenouille1956,Tukey1958}. This method is commonly used in statistics community to determine bias and variance (square of standard deviation) of a set of data points by using its subsets. We applied this method in quantifying the error in estimating magnetic exchange coupling parameters of bilinear (nearest neighbor)-biquadratic-three-site exchange model. 

The Jackknife estimate of a parameter is usually determined by estimating the parameter from each (n-1) subsample with taking out each $i^{th}$ data point from a given sample of size n.
The Jackknife estimate of mean is defined as,
\begin{equation}
\bar{J}=\frac{1}{n}\sum_{i=1}^{n}J_{i}.
\end{equation}
$J_{i}$ is estimated by solving system of (n-1) inhomogeneous linear equations when each $i^{th}$ equation is removed from system of n inhomogeneous linear equations.

The jackknife estimate of variance (square of standard error) is defined as,
\begin{equation}
var(J)=\frac{1}{n}\sum_{i=1}^{n}\bigg(\frac{J_{i}-\bar{J}}{\bar{J}}\bigg)^2.
\end{equation}

\begin{figure}
	\centering%
		\includegraphics[width=0.4\textwidth]{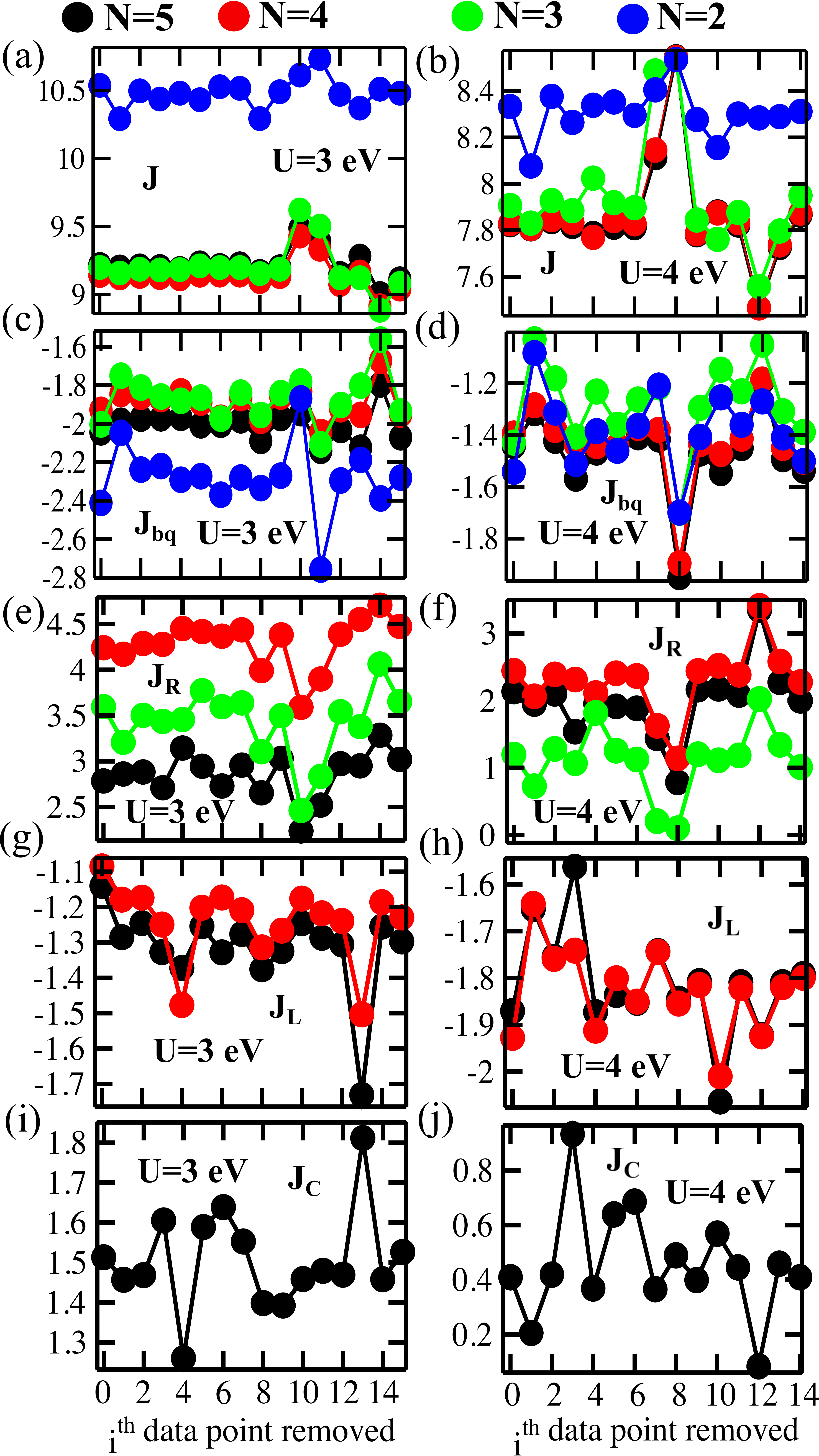}%
		\caption{Distributions of (a)-(b) $J$ (in meV), (c)-(d) $J_{bq}$ (in meV), (e)-(f) $J_R$ (in meV), (g)-(h) $J_C$ (in meV) and (i)-(j) $J_L$ (in meV) when estimated from system of n-1 linear equations (omitting each $i^{th}$ equation from system of n linear equations) for two different values of $U$ (=3 eV and 4 eV). N is the number of coupling parameters in the model Hamiltonian: two ($J$ and $J_{bq}$), three ($J$, $J_{bq}$ and $J_R$), four ($J$, $J_{bq}$, $J_R$ and $J_L$) and five ($J$, $J_{bq}$, $J_R$, $J_L$ and $J_C$).}%
	\label{FigS5}%
\end{figure}

The distributions of five coupling parameters ($J$, $J_{bq}$, $J_R$, $J_L$ and $J_C$) when fitted to bilinear (nearest neighbor)-biquadratic-three spin exchange model with two ($J$ and $J_{bq}$), three ($J$, $J_{bq}$ and $J_R$), four ($J$, $J_{bq}$, $J_R$ and $J_L$), five ($J$, $J_{bq}$, $J_R$, $J_L$ and $J_C$) parameters are presented in Fig. \ref{FigS5}. Distributions of $J$ (nearest neighbor bilinear coupling constant) and $J_{bq}$ (biquadratic coupling constant) estimated from model Hamiltonian with two parameters ($J$ and $J_{bq}$) are very different from that obtained using model Hamiltonian with three-site exchange (number of parameters more than two). Model Hamiltonian with three, four and five parameters (including three site exchange) give almost similar distributions of $J$ and $J_{bq}$ and highlights the necessity of using a model Hamiltonian with three site exchange parameters ($J_R$, $J_L$ and $J_C$).

\begin{figure}
	\centering%
		\includegraphics[width=0.8\textwidth]{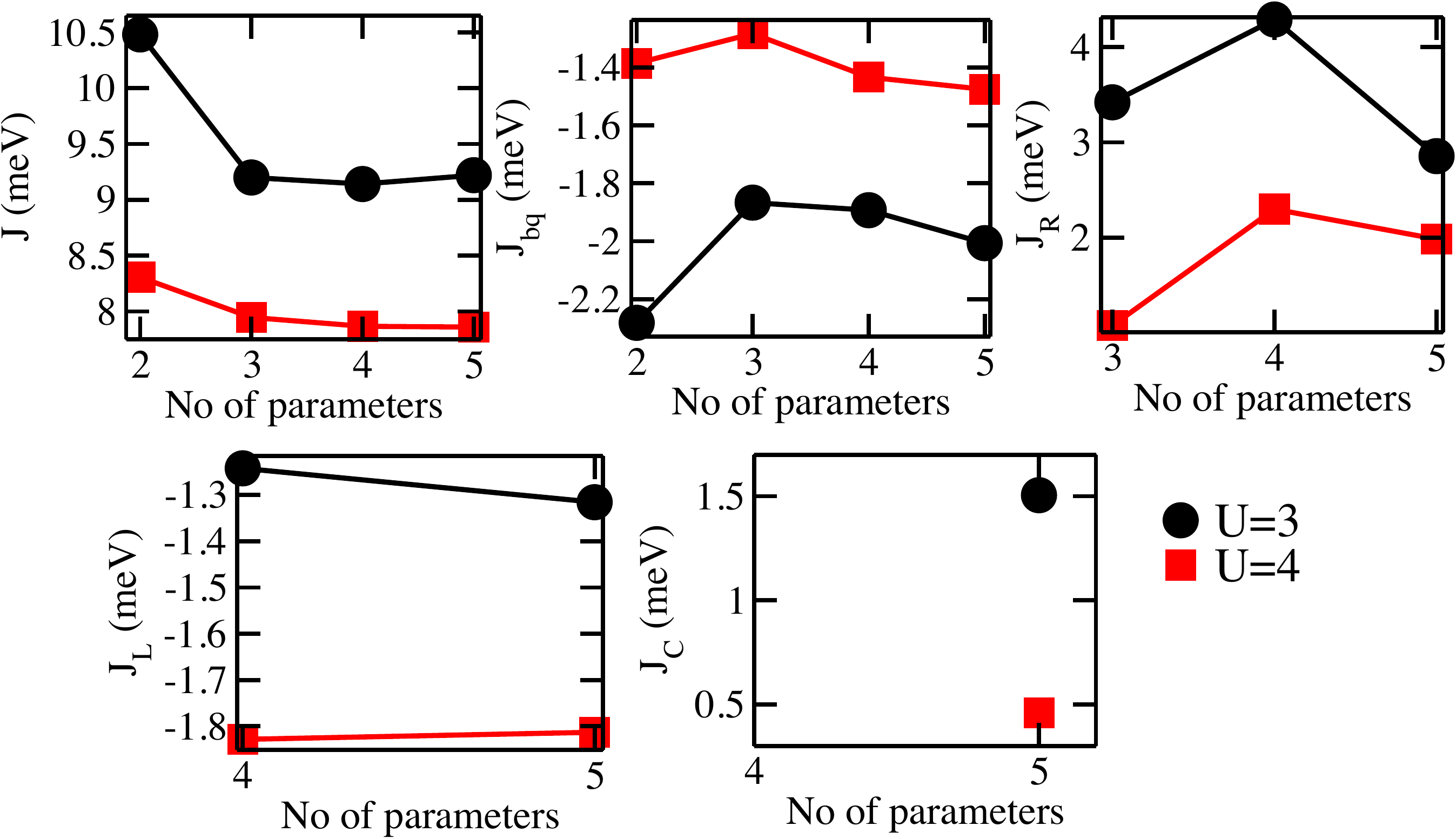}%
		\caption{Mean of distributions of $J$, $J_{bq}$, $J_R$, $J_C$ and $J_L$ as a function of number of coupling parameters in the model Hamiltonian: two ($J$ and $J_{bq}$), three ($J$, $J_{bq}$ and $J_R$), four ($J$, $J_{bq}$, $J_R$ and $J_L$) and five ($J$, $J_{bq}$, $J_R$, $J_L$ and $J_C$).}%
	\label{FigS6}%
\end{figure}

\begin{figure}
	\centering%
		\includegraphics[width=0.8\textwidth]{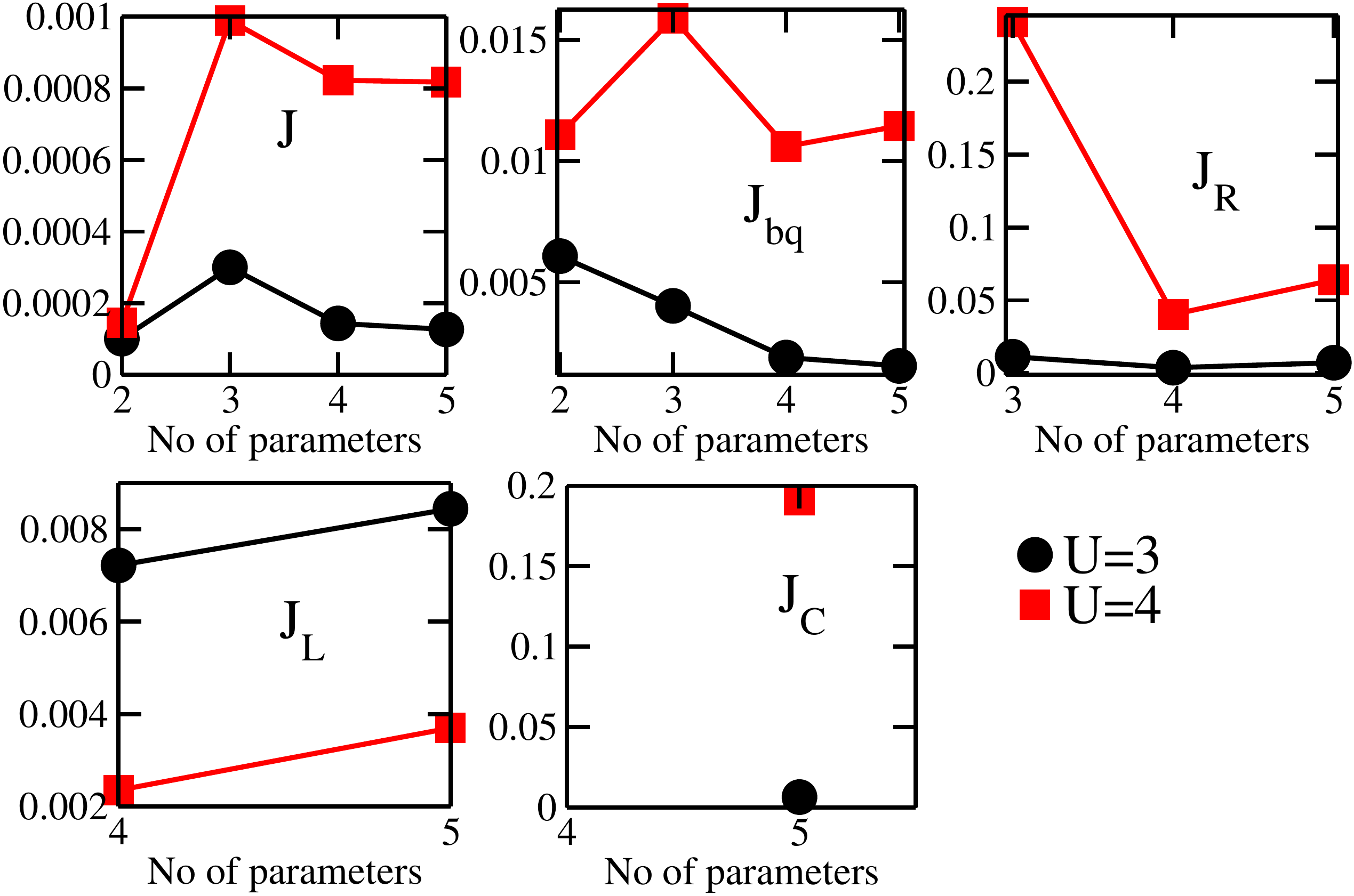}%
		\caption{Variance of distributions of $J$, $J_{bq}$, $J_R$, $J_C$ and $J_L$ as a function of number of coupling parameters in the model Hamiltonian: two ($J$ and $J_{bq}$), three ($J$, $J_{bq}$ and $J_R$), four ($J$, $J_{bq}$, $J_R$ and $J_L$) and five ($J$, $J_{bq}$, $J_R$, $J_L$ and $J_C$).}%
	\label{FigS7}%
\end{figure}

Mean of distributions of $J$ and $J_{bq}$ converges to certain value as the number of coupling parameters become larger than two and decreases as the $U$ parameter increases (see Fig. \ref{FigS6}). $J_R$ does not show any monotonic dependence on the number of coupling parameters as compared to $J$. $J_{bq}$. $J_L$ remains almost independent of number of the coupling parameters. 
The variance of distributions of all five coupling parameters is presented in Fig. \ref{FigS7}. The values of $J$, $J_{bq}$, $J_R$, $J_L$ and $J_C$ vary within 14\% ($U$=3 eV) from their mean values. In addition, $J_R$ and $J_C$ have slightly larger variance (within 40\% from mean) particularly for $U$=4 eV. Overall, $J_R$ and $J_C$ parameters do not change their sign as one of the data point is taken out for jackknife resampling (see Fig.~\ref{FigS5}). 

Two of the most important findings of the Jackknife analysis is that 1) The sign and order of magnitudes of all the parameters are the same for any sub-data set, or the Hamiltonian fitted. While the exact magnitude of various $J$'s depend on the model and the value of $U$ used in the DFT+U calculation, this makes our qualitative results robust. 2) A model with 3 parameters (i.e. one that includes $J_{R}$ reduces the variances of $J$, and $J_{bq}$. However, including other terms (such as $J_L$) does not do so. We thus conclude that a minimal model that explains the DFT data includes $J$, $J_{bq}$, and $J_R$. In principle, performing more DFT calculations with different spin configurations can increase the amount of input data to the fitting procedure, and thus can converge models with higher numbers of parameters, but as discussed in the main text, this is technically challenging. 

\clearpage

 \section{Structural distortion}

\begin{figure}[!h]%
	\centering%
		\includegraphics[width=0.6\textwidth]{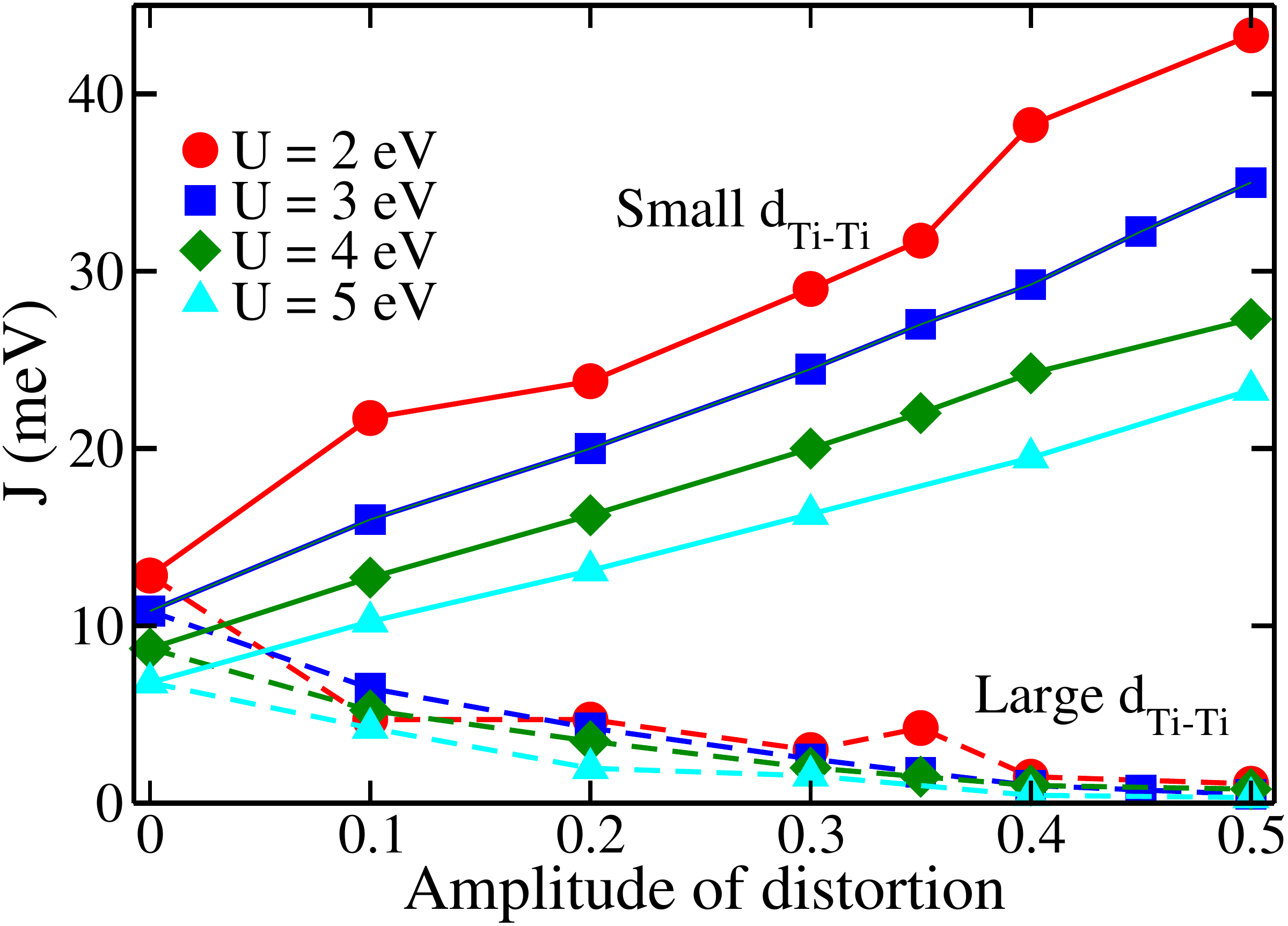}%
		\caption{Nearest neighbor bilinear coupling constant ($J$) as a function of structural distortion from the high temperature phase. The values of $J$ are determined by fitting energies of only collinear configurations to the model Hamiltonian with nearest neighbor bilinear coupling constant (fourth order coupling parameters $J_{bq}$, $J_R$, $J_C$ and $J_L$ are not considered in fitting).}%
	\label{FigS9}%
\end{figure}

To investigate the effect of displacements of Ti atoms from its position in the high temperature phase (ideal kagome lattice) on the nearest neighbor bilinear exchange ($J$), we extract the parameters for the magnetic Hamiltonian for the distorted crystal structures with the breathing Kagome structure. We repeat these calculations for various values of $U$ to make sure that the trends are robust, and we only consider the bilinear (Heisenberg) term for the nearest neighbors in the small triangles for simplicity. The structural distortion creates two inequivalent (one large and one small) Ti-Ti bond distances (see Fig. 1 in main manuscript). We find that the nearest neighbor bilinear coupling ($J$, see Fig. \ref{FigS9}) of spins separated by small distance increase nonlinearly as the increased orbital overlap leads to significant enhancement in the electron hopping. Decreasing the Ti-Ti bond length by 7\% (50\% distorted structure), nearest neighbor bilinear exchange increases tremendously by a factor of three from its value in the high temperature phase. In contrast, the bilinear coupling spins in the large triangle slowly goes to zero as the overlap of the orbitals decays exponentially. These findings suggest that the antiferromagnetic coupling of spins inside the small Ti triangle becomes stronger with distortion. The different choices of Hubbard $U$ give the similar trend in $J$. $J$ (bilinear coupling for small triangle) reduces with $U$ as the exchange interaction is inversely proportional to $U$.

\section{Wannier functions in the LT phase}

\begin{figure}
	\centering%
		\includegraphics[width=0.7\textwidth]{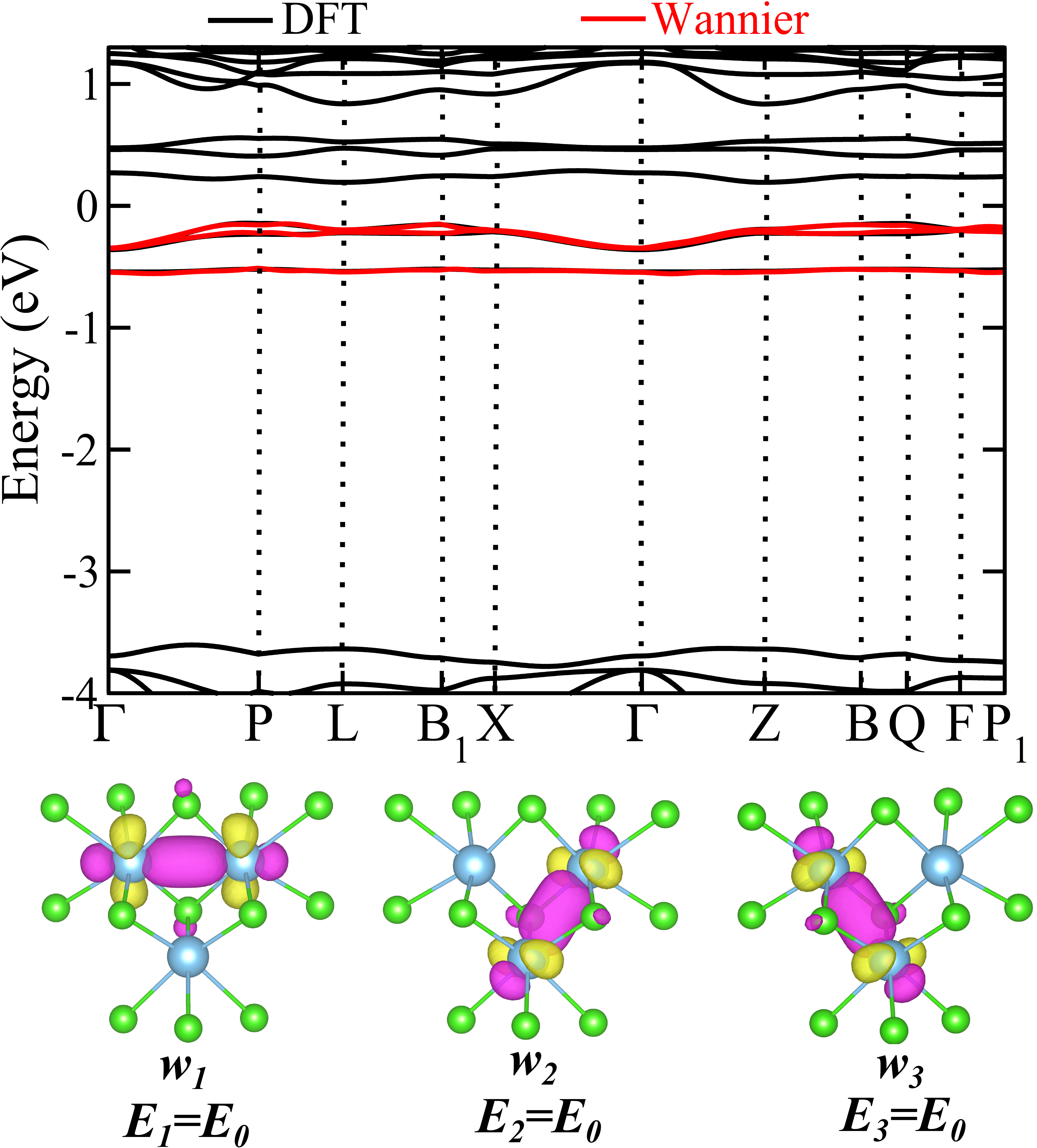}%
		\caption{Band structure calculated from DFT (black line) and Wannier interpolation (red colored bands derived from three Ti $d$ bands) in the low temperature phase. Lower panel shows bond centered Wannier functions. Light blue and green colored circles represent Ti and Cl atom respectively.}%
	\label{FigS8}%
\end{figure}

Wannier function calculations in the low temperature phase are performed without implementing any Hubbard $U$ parameter (DFT calculation). We consider three bands (just below the Fermi level) in the energy range from -0.21 eV to -0.55 eV, which are contributed by $d$ orbitals of Ti atom (see Fig. \ref{FigS8}). The maximally localized Wannier functions derived from these three bands (see Fig. \ref{FigS8}) are centered on Ti-Ti bonds. It originates from strong hybridization of $B_g$ orbitals (in the high temperature phase, see Fig.~3 
in the main manuscript) that leads to large inter-site hopping inside the small Ti triangle. 

\section{Phonons}

We determined zone centered phonon frequencies of Na$_2$Ti$_3$Cl$_8$ using frozen phonon method, and considered high temperature structure of Na$_2$Ti$_3$Cl$_8$ in the ferromagnetically and antiferromagnetically ordered and nonmagnetic state. The phonon mode with $\Gamma_2^-$ symmetry leads to the distortion of ideal kagome lattice (see Fig. 1 in main manuscript) and results in reduction of crystal symmetry from $R\bar{3}m$ (HT phase) to $R3m$ (LT phase). We find the high temperature structure to be dynamically stable (see TABLE. \ref{TABLE.I}) in both ferromagnetically and antiferromagnetically ordered state. However, a dramatic change in frequency of $\Gamma_2^-$ mode is observed as a function of Hubbard $U$. 
We find the ferromagnetic state to be metallic for small $U$ ($<$ 2 eV) whereas it is insulating for larger values of $U$ ($\geq$ 2 eV). We attribute this metal to insulator transition to the observed sudden jump in phonon frequency. We could only stabilize antiferromagnetic state for nonzero values of $U$ and this state remains insulating at all nonzero values of $U$. These findings suggest that one needs to go beyond the DFT method to capture the instability that leads to the trimerization of Ti atoms in the low temperature phase. In addition, the nonmagnetic state is metallic at all values of $U$. This state is energetically much higher ($>$ 1 eV/f.u.) compared to the magnetically ordered state and not dynamically stable (see TABLE \ref{TABLE.I}). 

\begin{table}[!htb]
\centering
\caption{Computed frequency (in cm$^{-1}$) of the $\Gamma_2^-$ phonon mode in the nonmagnetic (NM), ferromagnetic (FM) and antiferromagnetic (collinear AFM) states of Na$_2$Ti$_3$Cl$_8$. Text highlighted in yellow color is for the phonon frequency in the metallic state.}
\label{TABLE.I}
\begin{tabular}{cccc}
\noalign{\smallskip} \hline \hline \noalign{\smallskip}
Hubbard $U$ (eV) & NM & FM &  AFM\\
\hline
0.0 & \colorbox{yellow}{$i$386} & \colorbox{yellow}{40} & \\
1.0 & \colorbox{yellow}{$i$429} & \colorbox{yellow}{33} & 183 \\
2.0 & \colorbox{yellow}{$i$485} & 106 & 188 \\
3.0 & \colorbox{yellow}{$i$521} & 111 & 190 \\
4.0 & \colorbox{yellow}{$i$552} & 113 & 192 \\
\noalign{\smallskip} \hline \hline \noalign{\smallskip}
\end{tabular}
\end{table}

\section{DMRG and ED calculations for the $U=4$ eV DFT parameter set} 
In the main text we mentioned that our results and inferences about Na$_2$Ti$_3$Cl$_8$ 
are qualitatively robust to most choices of $U$ used in the functional. The DMRG and ED calculations shown corresponded 
to the parameters obtained from $U=3$ eV. Here we also show the results for the $U=4$ eV parameter set. 
%which corresponds to $J=7.8$ meV and $J_{R} \approx -J_{bq} = 1.5$ meV. 

\begin{figure*}
\includegraphics[width=0.47\linewidth]{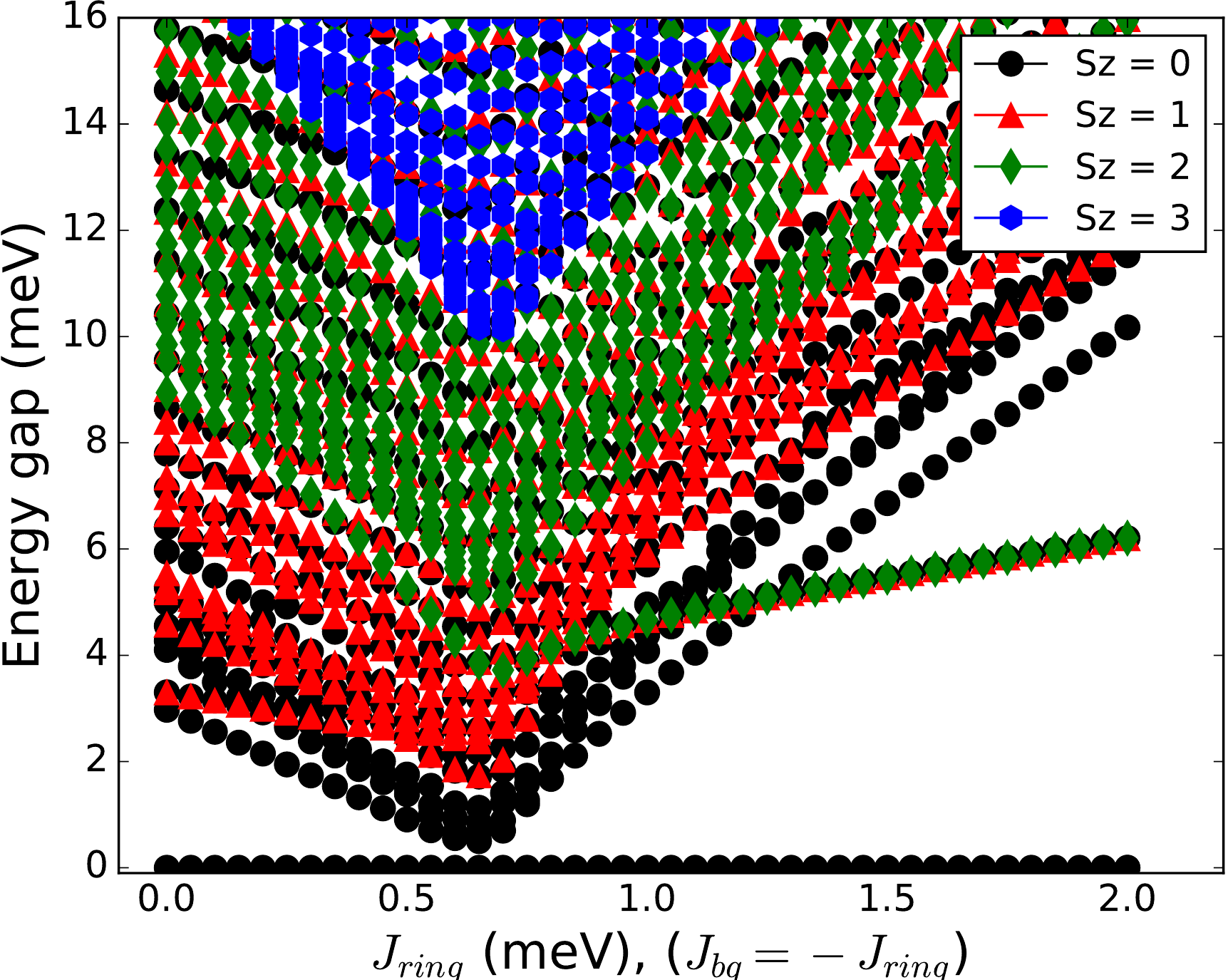}
\includegraphics[width=0.51\linewidth]{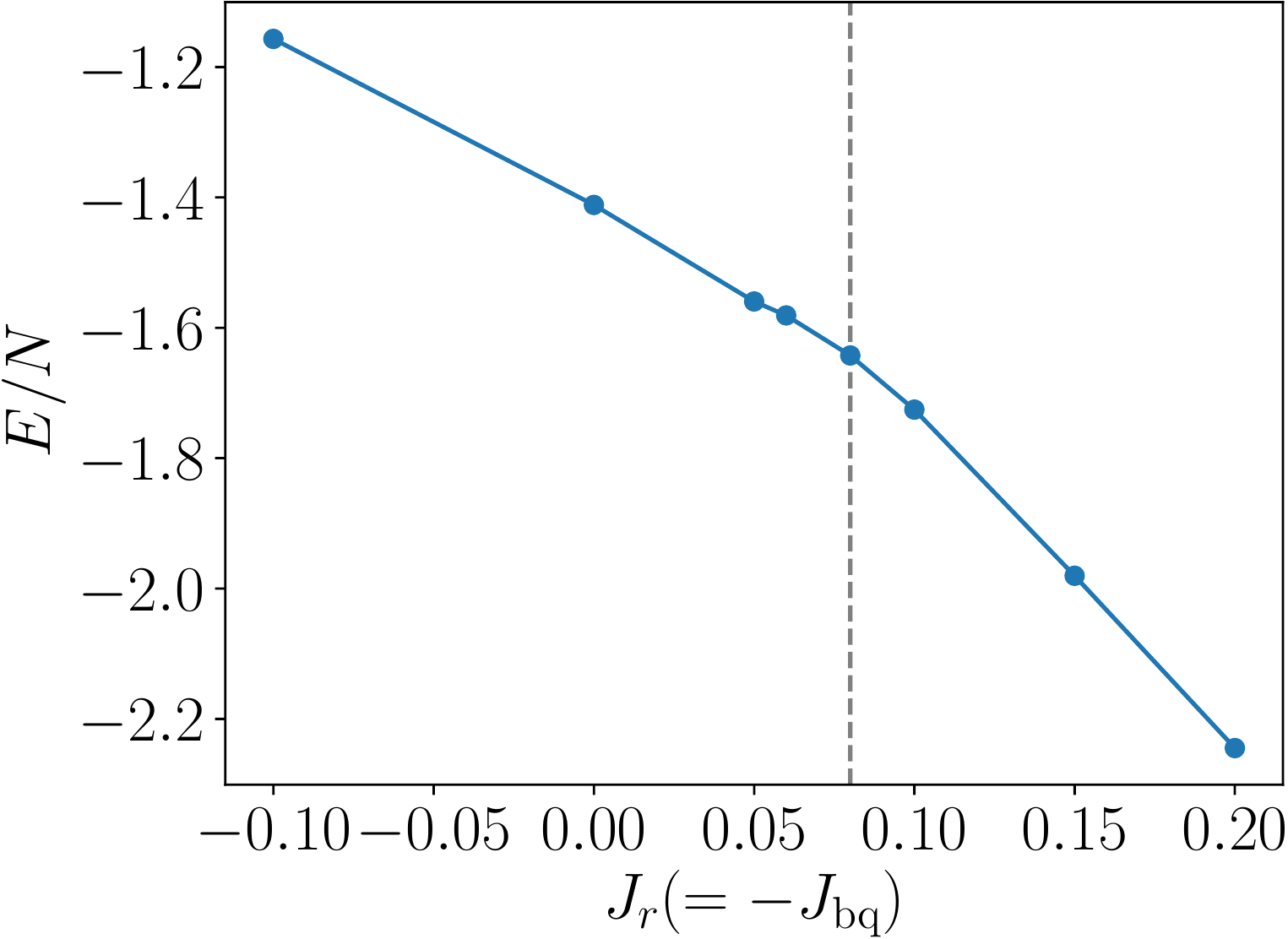}
\includegraphics[width=0.7\linewidth]{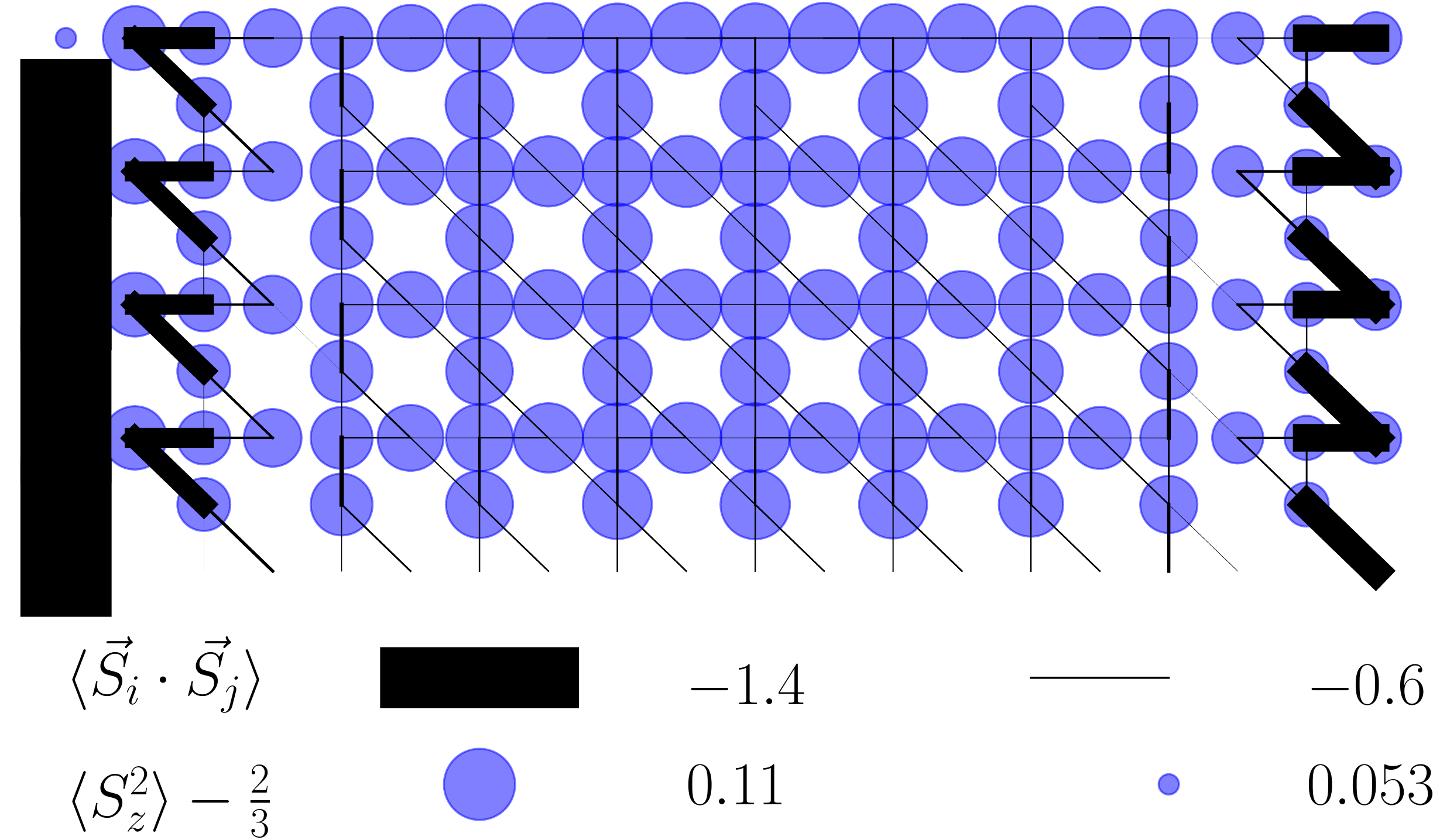}
\caption{(top left) Spectra from exact diagonalization on the 18b site cluster, for the $U=4$ parameter set, fixing $J=7.8$ meV and varying $J_{bq}=-J_{R}$. 
(top right) DMRG ground state energy (in units of $J=1$) as a function of $J_{R}/J$. 
(bottom) DMRG calculations on a kagome cylinder showing the nematic order parameter ($\langle (S^{z}_i)^{2} \rangle - \frac{2}{3} $) 
and bond expectation values $\langle \vec{S}_i \cdot \vec{S}_j \rangle$ for $J=1$,$J_{R}=-J_{bq}=0.2$.} 
\label{FigS10}%
\end{figure*}

Fig.~\ref{FigS10} (top left) shows the spectra (organized by $S_z$ sectors) from exact diagonalization on the 18b site cluster, 
for the $U=4$ parameter set, by fixing $J=7.8$ meV and varying $J_{bq}=-J_{R}$. 
The physical parameters correspond to $J_{R}=-J_{bq}=1.5$ meV, this suggests that 
the HT structure of the material is well in the nematic phase. A phase transition is seen at $J_{R}=-J_{bq} \approx 0.65$ meV 
corresponding to $J_{R}/J \approx 0.083$. In the panel on the top right, we plot the DMRG ground state energy 
(in units of $J=1$) as a function of $J_{R}/J$, which shows a kink at roughly the same location, confirming existence of a phase transition. 
In the bottom panel we explicitly confirm the presence of a nematic with 
DMRG calculations on a kagome cylinder that show that the nematic order parameter ($\langle (S^{z}_i)^{2} \rangle - \frac{2}{3} $) 
is non zero. 

\section{Origin of the biquadratic term in $S=1$ systems}
In the main text we discussed the importance of the biquadratic term that was needed to accurately fit the DFT energies. 
We found that in the high temperature structure, the value of the biquadratic coupling $J_{bq}$, 
could be 15 to 30 percent of the bilinear (Heisenberg) term $J$. %coupling to $\vec{S}_i \cdot \vec{S}_j$. 
Thus, it is important to understand if this scenario is physically plausible. 
Here we summarize briefly the microscopic origin of this term, motivated by the works of Bhatt and Yang (BY)~\cite{Bhatt_Yang} and Mila and Zhang (MZ)~\cite{Mila_Zhang}.

Both BY and MZ considered a lattice of atoms with two valence electrons per atom assuming the existence of a strong Hunds' coupling. 
For the present discussion, we summarize BY's argument adopting their notation and choice of Hamiltonian, however, the general result holds even in MZ's scheme. 
Similar to the case of Ti$^{2+}$ with a $t_{2g}$ manifold that is split, BY 
considered each ``site" to be made of $N$ atomic orbitals - the lowest energy level was two-fold degenerate (labelled by indices 1 and 2) with 
$N-2$ other nearly degenerate orbitals, whose energy is higher by an amount of $V$ (this is the crystal field splitting). 
Only intra-atomic interactions were considered, the energy cost for taking an electron from one site to the other to form a total spin of 3/2 
(parallel) and 1/2 (antiparallel) with the original spin-1 was denoted as $U_p$ and $U_a$ respectively. 
The hopping was considered to be,
\begin{equation}
	H_{hop} = - \sum_{a,b=1}^{N} t_{ab} \sum_{\sigma = \uparrow, \downarrow} \Big( c^{\dagger}_{ia\sigma}c_{jb\sigma} + \textrm{h.c.} \Big)
\end{equation} 
where $c_{ia\sigma}^{\dagger}$ is the creation operator of an electron in orbital $a$ on site $i$ with spin $\sigma$, and $t_{ab}$ is the 
hopping matrix element which is small compared to the scale of the Coulomb interactions i.e. the U's. To simplify the analysis, the model considered 
assumed only two independent parameters: $t_{ab}=t$ for $a=b$ and $t_{ab}=t$ for $a\neq b$. 

BY computed the energy change of the $S=0,1,2$ states in non-degenerate 
perturbation theory, on the basis of which they were able to obtain the effective couplings. They 
showed that to second order in perturbation theory, (ignoring the crystal field splitting), 
\begin{equation}
J^{(2)} = \frac{4 t'^2}{3} \Big(- \frac{N-2}{U_p} + \frac{N - 1/2} {U_a} \Big) + \frac{2 t^{2}}{U_a}   
\label{eq:J2}
\end{equation}
and the biquadratic coupling is \textit{exactly} zero. As expected, the sign of the Heisenberg coupling is positive, 
indicating antiferromagnetic Heisenberg interactions.

Proceeding to fourth order in perturbation theory, they showed, 
\begin{eqnarray}
J^{(4)} = - \frac{8 t'^{4}}{9 V} \Big( \frac{2}{{U_p}^2} - \frac{1}{{U_a}^2} + \frac{1}{U_a U_p}\Big) (N-2)^2     \\
J_{bq}^{(4)} = - \frac{2 t'^{4}}{9 V} \Big( \frac{4}{{U_p}^2} + \frac{1}{{U_a}^2} - \frac{4}{U_a U_p}\Big) (N-2)^2     
\end{eqnarray}
Note that the sign of the biquadratic coupling is negative, whereas the net Heisenberg coupling $J^{(2)} + J^{(4)}$ 
(for most physical reasonable parameters) will remain positive. This is broadly consistent the scenario encountered 
in our DFT fits. BY observed that $|J_{bq}|$ could be comparable to $J$ in the following situations 
(a) $V$ is small compared to $t$ (in which case the perturbative result 
does not strictly hold, but still expected to yield similar qualitative results in a modified formalism) (b) the orbital degeneracy $N$ is large (a situation that does not apply to Ti$^{2+}$ because $N=3$) (c) there is a cancellation of the terms of opposite signs in Eq.~\ref{eq:J2}

Finally, it must be emphasized that this entire analysis is based on some simplified assumptions about the hoppings and interactions, and thus it seems plausible 
that there could be models where a large $|J_{bq}/J|$ is possible. The objective here was to argue that simple models can also qualitatively capture 
the observations that we have quantitatively made with our numerical analysis of the DFT data.

\end{document}